# Statistical Models in Forensic Voice Comparison

Geoffrey Stewart Morrison [1,2]           https://orcid.org/0000-0001-8608-8207

Ewald Enzinger [1,3]                      https://orcid.org/0000-0003-2283-923X

Daniel Ramos [4]                          https://orcid.org/0000-0001-5998-1489

Joaquín González-Rodríguez [4]            https://orcid.org/0000-0003-0910-2575

Alicia Lozano-Díez [4]                    https://orcid.org/0000-0002-5918-8568

[1] Forensic Speech Science Laboratory, Aston Institute for Forensic Linguistics, and Forensic Data Science Laboratory, Computer Science Department, Aston University, Birmingham, United Kingdom

[2] Forensic Evaluation Ltd, Birmingham, United Kingdom

[3] Eduworks, Corvallis, Oregon, USA

[4] AUDIAS – Audio, Data Intelligence and Speech, Escuela Politécnica Superior, Universidad Autónoma de Madrid, Spain

**Abstract**

This chapter describes a number of signal-processing and statistical-modeling techniques that are commonly used to calculate likelihood ratios in human-supervised automatic approaches to forensic voice comparison. Techniques described include mel-frequency cepstral coefficients (MFCCs) feature extraction, Gaussian mixture model - universal background model (GMM-UBM) systems, i-vector - probabilistic linear discriminant analysis (i-vector PLDA) systems, deep neural network (DNN) based systems (including senone posterior i-vectors, bottleneck features, and embeddings / x-vectors), mismatch compensation, and score-to-likelihood-ratio conversion (aka calibration). Empirical validation of forensic-voice-comparison systems is also covered. The aim of the chapter is to bridge the gap between general introductions to forensic voice comparison and the highly technical automatic-speaker-recognition literature from which the signal-processing and statistical-modeling techniques are mostly drawn. Knowledge of the likelihood-ratio framework for the evaluation of forensic evidence is assumed. It is hoped that the material presented here will be of value to students of forensic voice comparison and to researchers interested in learning about statistical modeling techniques that could potentially also be applied to data from other branches of forensic science.

# 1   Introduction

The purpose of *forensic voice comparison* (aka forensic speaker comparison, forensic speaker recognition, and forensic speaker identification) is to assist a court of law to decide whether the voices on two (or more) recordings were produced by the same speaker or by different speakers. For simplicity, in the present chapter we will assume that there is one recording of a speaker of known identity (the *known-speaker recording*) and one recording of a speaker whose identity is in question (the *questioned-speaker recording*). Common scenarios include that the questioned-speaker recording is of a telephone call made to a call center, is of an intercepted telephone call, or is made using a covert recording device, and the known-speaker recording is of a police interview with a suspect or is of a telephone call made by a person who is in custody. The known-speaker recording is often an existing recording in which the identity of the speaker is not disputed, but sometimes a recording is made specifically for the purpose of conducting a forensic-voice-comparison analysis (practice varies depending on jurisdiction, laboratory policy, and the circumstances of the particular case).

There is usually a mismatch between the questioned- and known-speaker recordings in terms of speaker-intrinsic conditions, or speaker-extrinsic conditions, or both. Speaker-intrinsic variability can be due to multiple factors including differences in speaking style (e.g., casual versus formal and quiet versus loud), emotion (e.g., calm, angry, happy, sad), tiredness or physical stress (e.g., being out of breath), and elapsed time (the way a speaker speaks varies from minute to minute, hour to hour, day to day, etc. with larger differences occurring over longer time periods, Kelly & Hansen, 2016). In addition, the words and phrases that a speaker says vary from occasion to occasion. Speaker-extrinsic variability can be due to multiple factors including background noise that can vary in loudness and type (e.g., office noise, ventilation system noise, traffic noise, crowd noise – as well as noise being captured on the recording, speaking in a noisy environment also causes speakers to change the way they speak), reverberation (e.g., echoes in rooms with hard walls and floors), distance of the speaker from the microphone, the quality of the microphone and other components of the recording equipment, transmission of the recording through different communication channels that distort and degrade the signal in different ways (e.g., landline telephone, mobile telephone, Voice over Internet Protocol VoIP), and the format in which the recording is saved (in order to reduce the file size formats such as MP3 distort and degrade the signal). Intrinsic and extrinsic variability leads to mismatches between questioned- and known-speaker recordings within cases, and leads to different conditions and different mismatches from case to case.

Historically and in present practice, a number of different approaches have been used to extract information from voice recordings and a number of different frameworks have been used to draw inferences from that information. In *auditory* and *spectrographic* approaches information is extracted using subjective judgment, by listening to the recordings and by looking at graphical representations of parts of the recordings respectively (spectrograms are time by frequency by intensity plots of the acoustic signal). In conjunction with auditory and spectrographic approaches, inferences have almost invariably been drawn on the basis of subjective judgment. In *acoustic-phonetic* and *human-supervised automatic* approaches information is extracted in the form of quantitative measurements of the acoustic properties of the audio recordings. For the acoustic-phonetic approach, inferences can be drawn via statistical models, although in practice it is more common for practitioners of this approach to draw inferences on the basis of subjective judgment, e.g., by making a plot of the measured values from different recordings and then looking at the plot. For the automatic approach, inferences are invariably drawn via the use of statistical models. Even if inferences are drawn via statistical models, rather than directly reporting the output of the statistical model many practitioners use the output of the statistical model as input to a subjective judgment that also includes consideration of other information such as their subjective judgment based on auditory and acoustic-phonetic approaches (for arguments against this practice see Morrison and Stoel, 2014). It should be noted that even if the output of a quantitative-measurement and statistical-model approach is directly reported, such an approach still requires subjective judgments in decisions such as the choice of data used to train the statistical models. These subjective judgments are, however, as far removed as possible from the forensic practitioner's final conclusion, so of all the approaches this one is most resistant to cognitive bias (for recent reviews of cognitive bias in the context of forensic science see Found, 2015, Stoel et al., 2015, National Commission on Forensic Science, 2015, Edmond et al., 2017).

As of June 1, 2020, there are no published national or international standards specific to forensic voice comparison. The England and Wales Forensic Science Regulator's Codes of Practice and Conduct (Forensic Science Regulator, 2020) and their appendices relating to specific branches of forensic science are effectively national standards in that forensic laboratories in the UK can seek accreditation to the Codes, usually in combination with accreditation to ISO 17025:2015 General Requirements for The Competence of Testing and Calibration Laboratories. There is an appendix to the codes for Speech and Audio Forensic Services (Forensic Science Regulator, 2016). The European Network of Forensic Science Institutes (ENFSI) has published Methodological Guidelines for Best Practice in Forensic Semiautomatic and Automatic Speaker Recognition (Drygajlo et al., 2015). The Speaker Recognition Subcommittee of the Organization of Scientific Area Committees for Forensic Science (OSAC SR) is in the process of developing standards.

ENFSI guidelines include use of the likelihood-ratio framework for drawing inferences, and empirical validation of system performance under conditions reflecting those of the cases to which they are applied. The Forensic Science Regulator's codes also require methods to be validated. In forensic voice comparison these are not new ideas: quantitative-measurement and statistical-model based implementations of the likelihood-ratio framework date back to the 1990s, and calls for forensic-voice-comparison systems to be empirically validated under conditions reflecting casework conditions date back to the 1960s (for reviews see Morrison, 2009, and Morrison, 2014, respectively). The ENFSI guidelines also include the use of transparent and reproducible methods and procedures, and the use of procedures that reduce the potential for cognitive bias. Morrison and Thompson (2017) and Morrison (2018a) have argued that transparent quantitative-measurement and statistical-model based implementation of the likelihood-ratio framework with empirical validation under casework conditions would be the only practical way to comply with the admissibility criteria set out in United States Federal Rules of Evidence 702 and the *Daubert* trilogy of Supreme Court rulings (*Daubert v. Merrell Dow Pharmaceuticals*, 1993; *General Electric v. Joiner*, 1997; and *Kumho Tire v. Carmichael*, 1999), and those set out in England and Wales Criminal Practice Directions (2015) section 19A.

In order to reduce the potential for cognitive bias, increase transparency and reproducibility, facilitate validation under conditions reflecting casework conditions, and to draw inferences that are logically correct, we believe that the most practical approach is the *human-supervised automatic approach* used in conjunction with statistical-model based implementation of the *likelihood-ratio framework* with direct reporting of the output of the statistical model. The performance of acoustic-phonetic approaches have been found to be much poorer than the performance of automatic approaches (see Enzinger et al., 2012; Zhang et al., 2013; Enzinger, 2014; Enzinger and Kasess, 2014; Jessen et al., 2014; Enzinger and Morrison, 2017). Acoustic-phonetic approaches are also much more time-consuming and costly in skilled human labor, which makes empirical validation practically difficult. In the present chapter, we therefore discuss only the human-supervised automatic approach and statistical-model based implementation of the likelihood-ratio framework. We assume the reader is familiar with the likelihood-ratio framework, which has been described elsewhere in the present volume. Our aim is to provide an overview of a number of signal-processing and statistical-modeling techniques that are commonly used to calculate likelihood ratios in human-supervised automatic approaches to forensic voice comparison. We aim to bridge the gap between general introductions to forensic voice comparison and the highly technical (and often fragmented) automatic-speaker-recognition literature from which the signal-processing and statistical-modeling techniques are mostly drawn. The automatic-speaker-recognition literature is often fragmented because many influential papers are short conference-proceedings papers that do not provide fully detailed descriptions of the techniques they apply.

For readers unfamiliar with forensic voice comparison, we recommend general introductions such as Morrison and Thompson (2017), Morrison et al. (2018), and Morrison and Enzinger (2019). These include discussions of the likelihood-ratio framework, empirical validation, and legal admissibility. In the present chapter we go into greater technical detail regarding the calculation of likelihood ratios than is provided in such general introductions, but still attempt to make the material relatively accessible to an audience with a limited background in signal processing and statistical modeling. We have in mind researchers from other branches of forensic science and students of forensic voice comparison. There are alternatives to and multiple variants of many of the feature-extraction and statistical-modeling techniques we describe below. We do not attempt to be comprehensive, and describe only some of the variants that have commonly been used in forensic voice comparison. For overviews of automatic speaker recognition in general see Kinnunen and Lee (2010), Hansen and Hasan (2015), Fernández Gallardo (2016) §2.4, Ajili (2017) ch. 4, Matějka et al. (2020) (and see Hansen and Bořil, 2018, for a review of intrinsic and extrinsic variability in the context of automatic speaker recognition). These publications cover some of the same topics as the present chapter but are aimed at an audience with a background in signal processing and machine learning.

The chapter is structured as follows:

- Section 2 describes extraction of features from voice recordings, in particular extraction of mel-frequency cepstral coefficients (MFCCs).

- Section 3 describes mismatch compensation in the feature domain, in particular cepstral-mean subtraction (CMS), cepstral-mean-and-variance normalization (CMVN), and feature warping.

- Section 4 describes the Gaussian mixture model - universal background model (GMM-UBM) approach.

- Section 5 describes the identity-vector - probabilistic linear discriminant analysis (i-vector PLDA) approach, including mismatch compensation in the i-vector domain using canonical linear discriminant functions (CLDF). In the automatic-speaker-recognition literature, the latter is known as linear discriminant analysis (LDA).

- Section 6 describes deep neural network (DNN) based approaches, in particular those using senone posterior i-vectors, bottleneck features, and speaker embeddings (aka x-vectors).

- Section 7 describes score-to-likelihood-ratio conversion (aka calibration) using logistic regression. This can be applied to the output of GMM-UBM, i-vector PLDA, or DNN-based systems.

- Section 8 discusses empirical validation of forensic-voice-comparison systems.

GMM-UBM is an older approach, in use from about 2000. It has mostly been replaced by the i-vector PLDA approach, in use from about 2010. We describe GMM-UBM because it is still used by some forensic practitioners, it is somewhat more straightforward to describe than i-vector PLDA, and part of the discussion of the GMM-UBM approach provides a foundation for understanding the i-vector PLDA approach. Since about 2015 state-of-the-art systems in automatic-speaker-recognition research have been based on DNNs, and commercial DNN-based forensic-voice-comparison systems were first released around 2018. DNNs are generally used to create vectors (alternatives to i-vectors) that are then fed into PLDA models.

Color versions of the figures in this chapter and other material related to the chapter are available at http://handbook-of-forensic-statistics.forensic-voice-comparison.net/.

## 2    Feature extraction

In the context of forensic voice comparison, *features* are the acoustic measurements made on voice recordings. We describe the most commonly used features in automatic approaches to forensic voice comparison, mel-frequency cepstral coefficients (MFCCs), and their derivatives, deltas and double deltas.

For reviews of feature extraction methods in automatic speaker recognition, see Chaudhary et al. (2017),

Dişken et al. (2017), and Tirumala et al. (2017).

## 2.1   Mel-frequency cepstral coefficients (MFCCs)

*Mel-frequency cepstral coefficients* (MFCCs; see Davis and Mermelstein, 1980) are spectral measurements made at regular intervals, i.e., once every few milliseconds, during the sections of the recording corresponding to the speech of the speaker of interest. The noun "cepstrum" and adjective "cepstral" were coined by Bogert et al. (1963) via rearrangements of the letters in the words "spectrum" and "spectral". "Mel" refers to a frequency scaling that, unlike hertz, reflects human perception of frequency (Stevens et al., 1937). MFCCs are standard in speech processing in general, not just in automatic speaker recognition. In contrast to its use in automatic speech recognition, there is no principled reason for using mel scaling for automatic speaker recognition. Other variants of cepstral coefficients could be used, but MFCCs work well and are the most popular (Tirumala et al., 2017).

The steps for extracting MFCC measurements from voice recordings are described below, see also Figure 1 in which the numbers within black circles correspond to the numbered steps below.

1. The speech signal is multiplied by a bell-shaped window (e.g., a hamming window) with a duration typically on the order of 20 ms. The shape of the window is designed to reduce the impact of the window itself on the measurement of the spectrum (see Harris, 1978).

2. The power spectrum of the windowed signal is calculated using a discrete Fourier transform (DFT, or for computational efficiency a fast Fourier transform, FFT). A complex waveform, such as a windowed speech signal, can be constructed by adding together at each point in time the instantaneous intensities of a series of simple sine waves (the idea that any arbitrary waveform can be constructed using such a series is credited to Fourier, 1808). A Fourier analysis determines the frequencies, intensities, and phases of the sine waves that would have to be added together to make the observed complex waveform. The power spectrum consists of the frequencies and intensities of the sine wave components. For automatic speaker recognition, information about phase is usually discarded.

3. The power spectrum is multiplied by a filterbank. This is a series of triangular shaped filters, e.g., 26 filters, that are equally spaced on the mel-frequency scale. Each filter has a 50% overlap with each of its neighbors. Since forensic voice comparison often involves telephone recordings, the frequency range covered by the filters may be restricted to the traditional landline telephone bandpass (300 Hz – 3.4 kHz).

4. The intensities of the filterbank outputs are scaled logarithmically. The fact that this is log intensity will be relevant for feature domain mismatch compensation as described in Section 3 below.

5. A discrete cosine transform (DCT) is fitted to the output of the filterbank. A DCT is similar to a Fourier transform but all components are cosines of the same phase (or the opposite phase if the coefficient value is negative). The components are orthogonal: a constant, half a period of a cosine, one period of a cosine, one and a half periods of a cosine, etc. The process of fitting a DCT involves selecting values for the weights, or coefficients, on each component which result in the best fit to the data. Using all the DCT coefficients, the values of the original data can be recovered. Using only the first few DCT coefficients results in a smoothed version of the original data, i.e., a smoothed spectrum (aka a cepstrum). The higher order coefficients tend to capture statistical noise.

6. The first few DCT coefficients (e.g., the 1st through 14th DCT coefficients) are used as a vector of MFCCs. The 0th coefficient encodes the mean intensity of the signal and is usually discarded since this can be affected by factors not related to who is speaking (e.g., the distance of the speaker to the microphone or automatic gain control on the recording system).

The window is advanced in time, e.g., a 20-ms long window is advanced by 10 ms. Each range of time covered by a window is called a *frame*, and there is usually a 50% overlap between adjacent frames. Steps 1 through 6 are then repeated to produce another vector of MFCC values. The window is repeatedly advanced until MFCC vectors have been extracted from all sections of the recording corresponding to the speech of the speaker of interest. Since one MFCC vector is extracted every few milliseconds, a relatively large number of feature vectors is extracted, e.g., 100 feature vectors per second if the frame advance is 10 ms.

<Insert Figure 1 about here>

## 2.2    Deltas and double deltas

*Deltas* are derivatives of MFCCs and encode the local rate of change of MFCC values over time (see Furui, 1986). *Double deltas* are the second derivatives of MFCCs and encode the local rate of change of delta values over time. Vectors of deltas and double deltas are concatenated with the MFCC vectors to produce longer *feature vectors*, e.g., if the original MFCC vector has 14 values, the MFCC + delta vector will have 28, and the MFCC + delta + double delta vector will have 42.

Delta values are calculated separately for each MFCC dimension. In each dimension, a linear regression is fitted to a contiguous set of MFCC values, e.g., the MFCC value at the frame in time where the delta is being measured plus the MFCC values from the same dimension in the two preceding and the two following frames. Measurements are usually made over the ±2 or ±3 adjacent frames. The value of the slope of the fitted linear regression is used as the delta value.

Double deltas are calculated in the same way as deltas, but based on the delta values rather than on the MFCC values.

The statistical modeling steps of automatic-speaker-recognition systems usually discard information regarding the original time sequence of the feature vectors, hence the deltas and double deltas encode the only time-sequence information that is exploited. DNN embedding systems are an exception to this.

## 2.3    Voice-activity detection (VAD) and diarization

Either prior to or after extracting MFCCs, the speech of the speaker of interest should be separated from periods of silence, transient noises, and the speech of other speakers. Either only the speech of the speaker of interest should be measured, or only measurements corresponding to the speech of the speaker of interest should be kept. This is done by either manually or automatically marking the beginning and end of each utterance (stretch of speech) of interest. An automatic procedure may be followed by manual checking and correction as needed.

The process of finding utterances is called *voice-activity detection* (VAD, aka speech-activity detection, SAD). A simple automatic *voice-activity detector* (also abbreviated as VAD) may be based solely on root-mean-square (RMS) amplitude, and thus simply find louder parts of the recording. Such simple

VADs may not, however, perform well under conditions that include background noise, as is common in forensic voice comparison (see Mandasari et al. 2012). More sophisticated VADs employ algorithms to distinguish speech from other sounds and noises (e.g., Sohn et al., 1999; Beritelli and Spadaccini, 2011; Sadjadi and Hansen, 2013).

The process of attributing different utterances in the recording to different speakers is called *diarization*. Automatic diarization is itself a form of automatic speaker recognition and may itself make use of MFCCs.

## 3 Mismatch compensation in the feature domain

The acoustic properties of voice recordings can vary because they are produced by different speakers, but also because of other factors such as differences in speaking styles (e.g., casual, formal, whispering, shouting), differences in background noise and/or reverberation, differences in the distance of the speaker to the microphone, different types of microphones, transmission through different communication channels (e.g., landline telephone, mobile telephone, VoIP), and being saved in different formats (in order to reduce the size of the files stored, formats such as MP3 use lossy compression which discards some acoustic information and distorts remaining acoustic information). Mismatch compensation techniques seek to maximize between-speaker differences and minimize differences due to other factors. Feature domain mismatch compensation techniques generally attempt to reduce differences due to recording and transmission channels and differences due to acoustic noise.

We describe three commonly used feature domain mismatch compensation techniques: cepstral-mean subtraction (CMS), cepstral-mean-and-variance normalization (CMVN), and feature warping.

### 3.1 Cepstral-mean subtraction (CMS) and Cepstral-mean-and-variance normalization (CMVN)

*Cepstral-mean subtraction* (CMS; Furui, 1981) as a mismatch compensation technique is based on the premise that the speech signal, which is changing rapidly over time, is convolved with a channel that is invariant over time (it is linear time invariant, LTI). This is a good model for a traditional landline telephone system, the effect of which is essentially to pass the speech signal through a bandpass filter. Different microphones have different frequency responses. Some may be more sensitive to lower frequency sounds, some to higher frequency sounds, etc. Microphones and other basic components of recording systems can be treated as linear time invariant. Differences in the signal due to differences in the distance from the speaker to the microphone can also be treated as linear time invariant effects, as long as the distance does not change during the recording.

Convolution in the time domain is equivalent to multiplication in the frequency domain. Since MFCCs are frequency domain representations they can be considered the result of multiplying the dynamic speech signal and the invariant channel. But note that in generating MFCCs, intensity was logarithmically scaled (Step 4 in Section 2.1 above). Multiplication on a linear scale is equivalent to addition on a logarithmic scale, hence we should consider MFCCs the result of adding the speech signal and the channel in the log frequency domain.

Assuming the channel is invariant over time, it can be estimated by taking the mean of the cepstral coefficients over time. For the value in each dimension in each frame, the mean for that dimension is subtracted. What remains are the dynamic aspects of the original MFCC features. Deltas and double deltas are calculated on the raw MFCCs before CMS is applied to the MFCCs, and (although not obviously theoretically motivated) CMS is then usually applied to the deltas and double deltas.

As well as removing the effect of the invariant channel, CMS also removes the speech signal's mean. If there is a substantial channel mismatch between the questioned- and known-speaker recordings, removing the effect of the channel will tend to lead to better performance despite the loss of the speech-signal mean information (but CMS will tend to lead to worse performance if there is in fact no channel mismatch, Reynolds, 1994). For each recording, the mean of the feature values in each dimension will be 0, hence statistical models applied to the post-CMS features are designed to exploit what remains: non-normalities (in terms of skewness, kurtosis, and multimodalities), variance, and multidimensional correlations.

Although the theoretical motivation is less immediately obvious, *cepstral-mean-and-variance normalization* (CMVN; Vikki and Laurila, 1998) is a statistically obvious extension to CMS in which, as well as subtracting the mean, the variance of each MFCC dimension is scaled to 1.

CMS and CMVN can be applied globally, i.e., using the mean and variance of MFCC values from the whole of the speech of the speaker of interest in a particular recording, or can be applied locally, i.e., using the mean and variance of the MFCC values from the speech of the speaker of interest within a window that is a few seconds long. Such local application is discussed below in the context of feature warping.

## 3.2    Feature warping

Rather than assuming that there is a channel effect that is invariant throughout the recording, *feature warping* (Pelecanos and Sridharan, 2001) is based on the assumption that there may be a slowly changing channel effect convolved with a rapidly changing speech signal. It also assumes that there may be slowly changing additive noise. Since these are assumed to be slowly changing, their effect on the MFCC values is assumed to be stable locally, i.e., over a period of a few seconds.

The implementation of feature warping is described below, see also Figure 2. Feature warping is applied separately to each feature vector dimension. It is based on finding where the current feature value lies relative to the empirical distribution of the values in, for example, the preceding 1.5 s and the following 1.5 s (±150 frames), and then warping that value to the corresponding value in a target distribution. The target distribution is usually a Gaussian distribution with a mean of 0 and variance of 1.

1. Obtain the feature values from the 150 frames preceding the current frame, and the 150 frames following the current frame.

2. Sort the 301 values (including the current value) in ascending order, and calculate each value's rank proportional to the total number of values. This can be represented graphically as a plot of the empirical cumulative distribution of the feature values.

3. Find the value of the empirical cumulative distribution, $y_{\text{empirical}}$, which corresponds to the current feature value, $x_{\text{original}}$.

4. On the target cumulative probability distribution, locate the point $y_{\text{target}} = y_{\text{empirical}}$.

5. Read off the corresponding $x_{\text{warped}}$ value from the target cumulative probability distribution.

<Insert Figure 2 about here>

Deltas and double deltas are calculated on the raw MFCCs before feature warping is applied to the MFCCs, and feature warping is then applied to the deltas and double deltas.

Figure 3 shows an example of the effect of applying feature warping to mismatched recording conditions. The histograms in the leftmost column represent the distribution of the 1st order MFCC values extracted from a studio-quality speech recording. The histograms in the next three columns show the distributions of the 1st order MFCC values extracted after having applied various signal processing techniques to simulate the different conditions of a questioned-speaker recording and a known-speaker recording from a case (top and bottom rows respectively). Notice how the distributions change after each step and how the distributions under the simulated questioned- and known-speaker conditions diverge from one another. The histograms in the rightmost column show the distributions after feature warping is applied. Notice how the distributions are now warped to approximate the same target distribution.

<Insert Figure 3 about here>

Making all univariate distributions the same may seem extreme, but the ordering within vectors is not altered and so correlations across dimensions are not lost. The correlations of interest are ultimately due to time and frequency patterns in the acoustic signal resulting from the articulation of speech sounds. Particular time and frequency patterns result from the articulation of particular speech sounds. Statistical models applied to the warped features are designed to exploit these multidimensional correlations. Empirically, feature warping can outperform CMS and CMVN (see for example: Pelecanos and Sridharan, 2001; Silva and Medina, 2017).

## 4 GMM-UBM

This section describes the *Gaussian mixture model - universal background model* approach (GMM-UBM; see Reynolds et al., 2000).

The GMM-UBM model is a *specific-source model*,[1] i.e., it builds a model for the specific known-speaker recording in the case, and thus answers the following two-part question:

---

[1] For a discussion of the distinction between *specific-source* and *common-source* models, see Ommen and Saunders (2018).

> What would be the probability of obtaining the feature vectors of the questioned-speaker recording if the questioned speaker were the known speaker?
>
> versus
>
> What would be the probability of obtaining the feature vectors of the questioned-speaker recording if the questioned speaker were <u>not</u> the known speaker, but some other speaker selected at random from the relevant population?

Conceptually, the answer to the first part of the question provides the numerator for the likelihood ratio and the answer to the second part provides the denominator.

A *Gaussian mixture model* (GMM), is a probability density function constructed by adding together a number of individual Gaussian distributions. Assuming a multivariate distribution, each of the $G$ component Gaussian distributions, $f(x|\boldsymbol{\mu}_g, \boldsymbol{\Sigma}_g)$, in the mixture has a mean vector $\boldsymbol{\mu}_g$, a covariance matrix $\boldsymbol{\Sigma}_g$, and a weight $w_g$, see Equation 1. Usually, diagonal-only covariance matrices are used – a larger number of Gaussian components with diagonal-only covariance matrices can achieve the same degree of fit as a smaller number of components with full covariance matrices, and the former approach is more computationally efficient. In the present context, $x$ is a feature vector after application of feature domain mismatch compensation, and $y$ is the probability density (likelihood) of the model evaluated at $x$.

(1)
$$y = \sum_{g=1}^{G} w_g f(x|\boldsymbol{\mu}_g, \boldsymbol{\Sigma}_g), \quad \sum_{g=1}^{G} w_g = 1, \quad w_g \geq 0$$

## 4.1 Training the relevant-population model (the UBM): Expectation maximization (EM) algorithm

The mean vectors, covariance matrices, and weights for a GMM are trained using the *expectation maximization* (EM) algorithm (Dempster et al., 1977; Hastie et al., 2009, §8.5). Although the data in forensic voice comparison are multivariate, for simplicity of exposition we describe the EM algorithm below assuming univariate data. Also see Figure 4 which shows an example of a two-component GMM being trained on artificial data that were created for illustrative purposes. In the one dimensional case, assume that we have scalar training data, $x_i, i \in \{1, \ldots, N\}$.

1. The number of Gaussian components, $G$, to use must be specified. A commonly used number of Gaussian components is 1024 (the optimal value will depend on the particular application and amount of training data available).

2. A starting point is needed. A common approach is to use a k-means clustering algorithm (see Arthur and Vassilvitskii, 2007; Hastie et al., 2009, §14.3) to cluster the data into the same number of clusters as there are Gaussian components. For each Gaussian component, the initial mean value, $\mu_{g,0}$, is trained using the data from one of these clusters. The initial variance, $\sigma_{g,0}^2$, for all components is usually set to the sample variance of the whole of the training data (in the multivariate case the initial covariance matrix is diagonal only). Equal initial weights, $w_{g,0} = 1/G$, are usually assigned to all components.[2]

3. The *expectation step* asks how likely each training datum is assuming it had it come from the distribution of one of the Gaussian components. For each training datum, $x_i$, this question is repeated for each of the $G$ components. The answers, $\gamma_{g,i}$, are called *responsibilities*. Each is calculated as the relative likelihood for the combination of a particular training datum, $x_i$, and a particular Gaussian component, $g$, see Equation 2.

(2)

$$\gamma_{g,i} = \frac{w_{g,0} f(x_i | \mu_{g,0}, \sigma_{g,0})}{\sum_{j=1}^{G} w_{j,0} f(x_i | \mu_{j,0}, \sigma_{j,0})}$$

---

[2] Other approaches include binary splitting (see Ueda et al., 2000), and random seed. For the latter, the initial mean value, $\mu_{g,0}$, of each component is set to a randomly selected $x_i$ value from the training data. If a random seed is used, each random seed will lead to a different solution. Training may be repeated multiple times using multiple random seeds and the result with the best fit to the data used.

4. The *maximization step* recalculates the means, variances, and weights. To calculate the new mean, $\mu_{g,1}$, and new variance, $\sigma_{g,1}^2$, for a Gaussian component, $g$, all the training data are used with each datum, $x_i$, weighted by its responsibility, $\gamma_{g,i}$, see Equation 3 and 4. The new weight for the Gaussian component, $w_{g,1}$, is the mean of the responsibilities for that component, see Equation 5.

(3)
$$\mu_{g,1} = \frac{\sum_{i=1}^{N} \gamma_{g,i} x_i}{\sum_{i=1}^{N} \gamma_{g,i}}$$

(4)
$$\sigma_{g,1}^2 = \frac{\sum_{i=1}^{N} \gamma_{g,i} (x_i - \mu_{g,1})^2}{\sum_{i=1}^{N} \gamma_{g,i}}$$

(5)
$$w_{g,1} = \frac{1}{N} \sum_{i=1}^{N} \gamma_{g,i}$$

5. Steps 3 and 4 are repeated multiple times, each time updating the means, variances, and weights for all $G$ Gaussian components. Each repetition is called an *iteration*.

6. The algorithm stops when it converges on a solution. Convergence occurs when, from one iteration to the next, the change in the goodness of fit of the model to the training data becomes smaller than a pre-specified threshold. Alternatively, the algorithm stops after a pre-specified number of iterations.

<Insert Figure 4 about here>

The first step in using a GMM-UBM is to train a GMM using feature vectors extracted from recordings of a sample of speakers representative of the relevant population for the case. Feature vectors from recordings of all the speakers in the sample are pooled and used to train a GMM which is called a *universal background model* (UBM). This is the model which will be used to calculate the denominator of the likelihood ratio.

## 4.2 Training the known-speaker model: Maximum a posteriori (MAP) adaptation

The model for calculating the numerator of the likelihood ratio, a *speaker model*, is also a GMM, but rather than training the model from scratch, the known-speaker GMM is adapted from the UBM (Reynolds et al., 2000). One reason for this is that successful training of a GMM with a large number of Gaussian components in a high dimensional space requires a large amount of training data. The UBM is trained using data from a large number of speakers, but the known-speaker GMM has to be trained using data from one speaker, and often only one relatively short recording of that speaker is available. The procedure for adapting a single-speaker GMM from the UBM is a form of *maximum a posteriori adaptation* (MAP). It is similar to the EM algorithm, but usually only one iteration of MAP adaptation is applied, usually only the means are adapted, and the means are only partially adapted. A new mean is the result of a weighted mixture of the original mean from the UBM, $\mu_{g,\text{UBM}}$, and what would be the new mean if the standard EM algorithm were applied, $\mu_{g,\text{EM}}$ (the latter is $\mu_{g,1}$ in Equation 6a above). The new mean is calculated using Equation 6, in which $\alpha_g$, the weight for mixing the EM and UBM means, is known as the *adaptation coefficient*, and $\tau$ is known as the *relevance factor*.

(6)

$$\mu_{g,1} = \alpha_g \mu_{g,\text{EM}} + (1 - \alpha_g)\mu_{g,\text{UBM}} = \alpha_g \frac{\sum_{i=1}^{N} \gamma_{g,i} x_i}{\sum_{i=1}^{N} \gamma_{g,i}} + (1 - \alpha_g)\mu_{g,0}$$

$$\alpha_g = \frac{N w_{g,1}}{N w_{g,1} + \tau}$$

If the new weight for a Gaussian component, $w_{g,1}$, is large, i.e., lots of adaptation training data are associated with that Gaussian component, then $\alpha_g$ is larger and the new mean depends more on the EM mean, whereas if $w_{g,1}$ is small, $\alpha_g$ is smaller and the new mean depends more on the UBM mean.[3] This can therefore be thought of as a form of Bayesian adaptation with the UBM mean as the prior mean. The more sample data associated with a Gaussian component the closer its posterior mean will be to the sample mean. Increasing $\tau$ gives globally greater weight to the UBM means (the value for $\tau$ used in Reynolds et al., 2000, was 16).

Figure 5 shows an example of a two-dimensional UBM and a MAP adapted known-speaker GMM.

<Insert Figure 5 about here>

### 4.3    Calculating a score

Assume that a multivariate UBM population model was trained on a sample of data from the relevant population, and has mean vectors $\boldsymbol{\mu}_{r_j}$, covariance matrices $\boldsymbol{\Sigma}_{r_j}$, and weights $w_{r_j}$, $j \in \{1 \ldots G\}$. Assume that a multivariate GMM known-speaker model was trained on data from the known speaker, and has mean vectors $\boldsymbol{\mu}_{k_j}$, covariance matrices $\boldsymbol{\Sigma}_{k_j}$, and weights $w_{k_j}$, $j \in \{1 \ldots G\}$. Also, assume that the data from the questioned-speaker recording consists of $N_q$ feature vectors: $\boldsymbol{x}_{q_i}$, $i \in \{1 \ldots N_q\}$.

To calculate a likelihood ratio, $\Lambda_{q_i,k}$, for a single feature vector from the questioned-speaker recording, $\boldsymbol{x}_{q_i}$, the likelihood of the known-speaker model is evaluated given that feature vector, the likelihood of the population model is evaluated given that feature vector, and the former is divided by the latter, see Equation 7 and a graphical example in Figure 5.

(7)

$$\Lambda_{q_i,k} = \frac{\sum_{j=1}^{G} w_{k_j} f\left(\boldsymbol{x}_{q_i} \middle| \boldsymbol{\mu}_{k_j}, \boldsymbol{\Sigma}_{k_j}\right)}{\sum_{j=1}^{G} w_{r_j} f\left(\boldsymbol{x}_{q_i} \middle| \boldsymbol{\mu}_{r_j}, \boldsymbol{\Sigma}_{r_j}\right)}$$

---

[3] If only the means are adapted, the "new" weights are only used for this calculation, and it is actually the old UBM weights that are used for the speaker model.

Note, however, that if a feature vector is extracted every 10 ms, then there will be 100 feature vectors for every second of speech in the questioned-speaker recording. The total number of feature vectors, $N_q$, from the questioned speaker recording may be in the thousands or tens of thousands. We do not want to report a likelihood ratio at the first 10-ms mark, a likelihood ratio at the second 10-ms mark, etc. We want to report a single value quantifying the strength of evidence associated with the whole of the questioned-speaker speech. Our next step toward this is to calculate the mean of the per-feature-vector log likelihood ratios, as in Equation 8.

(8)
$$S_{q,k} = \frac{1}{N_q} \sum_{i=1}^{N_q} \log(\Lambda_{q_i,k})$$

We will call the mean of the per-feature-vector log likelihood ratios, $S_{q,k}$, a *score*. We will not call it a likelihood ratio. Multiplying all the per-feature-vector likelihood ratios together, or adding all the per-feature-vector log likelihood ratios together, is known as naïve Bayes fusion. It is naïve because it assumes there is no correlation between the likelihood-ratio values being combined. So that score values are not heavily dependent on the duration of the questioned-speaker recording, a score is calculated as the mean of the per-feature-vector log likelihood-ratio values rather than as their sum, but this still ignores correlation. In fact, the likelihood-ratio values come from a series of feature vectors taken from adjacent frames of speech recording with 50% overlap between adjacent frames (see Section 2.1). Substantial correlation between the frame-by-frame likelihood-ratio values is therefore expected. In addition, in training the UBM and the GMM speaker models we have estimated a large number of parameter values, e.g., 42 dimensions × (1024 means + 1024 variances) + 1024−1 weights = 87,039 parameter values. Unless we had an extremely large amount of data, those estimates may be poor. Thus, we are not safe to treat the value of $S_{q,k}$ as an appropriate answer to the question posed by the same-speaker and different-speaker hypotheses in the case. To fix this problem, we will implement an additional step: score-to-likelihood-ratio conversion (aka calibration). This is the topic of Section 7 below – some readers may wish to skip directly to Section 7 and return to Sections 5 and 6 later.

### 4.4    Remarks regarding UBM training data

The UBM is the model which represents the relevant population. The data used to train the UBM should therefore be a sample that is representative of the relevant population in the case. In addition, the training data should reflect the speaking style and recording conditions of the known-speaker recording. Any mismatch between the questioned-speaker data and the data used to train the known-speaker model, the model in the numerator of the likelihood ratio, will then be the same as the mismatch between the questioned-speaker data and the data used to train the population model, the model in the denominator of the likelihood ratio. Feature-level mismatch compensation techniques cannot be assumed to be 100% effective, and a difference in the conditions of the data used to train the model in the numerator and the model in the denominator would be expected to bias the calculated value of the score (see Morrison, 2018b).

## 5    i-vector PLDA

### 5.1    i-vectors

An *i-vector* is a single vector representing the speaker information extracted from a single recording. The "i" stands for "identity". The lengths of i-vectors extracted from different recordings are the same irrespective of the lengths of the recordings. i-vectors are described in: Kenny et al. (2005); Matrouf et al. (2007); Dehak et al. (2011); Matějka et al. (2011); Bousquet et al. (2013).

Below we describe a set of procedures for calculating i-vectors that are one of the sets of procedures tested in Bousquet et al. (2013). These are not the most commonly used procedures, but are approximately equivalent to the more commonly used procedures and easier to explain and understand. Further below, we give a brief overview of the more commonly used procedures, with additional details provided in Appendix A.

To generate an i-vector from a speech recording, we begin by training a UBM on feature vectors extracted from recordings of a large number of speakers under a variety of recording conditions. The standard approach is to use a very large diverse set of speakers in a diverse range of recording conditions. Ideally, the training data should include multiple recordings from each speaker, the multiple recordings including different recording conditions. In the standard approach the training data for the UBM do not represent the case-specific relevant population or the case-specific conditions.[4]

After training the UBM, the next step is to train a GMM for each recording in a set of recordings that are representative of the relevant population for the case and that reflect the recording conditions of the questioned- and known-speaker recordings in the case (when the amount of case-relevant data is small, domain adaptation may be used, see García-Romero and McCree, 2014). The GMM for each recording is trained using mean-only MAP adaptation from the UBM. Since each GMM is adapted from the same UBM, the mean vectors in different GMMs have a parallel structure (they lie in the same vector space), e.g., the first mean of the mean vector of the first component in the GMM for one recording is parallel to the first mean of the mean vector of the first component in the GMM for another recording because they were both adapted from the first mean of the mean vector of the first component in the UBM, *mutatis mutandis* for every mean in the mean vector of every component. We take all the mean vectors from all the components in the GMM, and concatenate them to form a *supervector*. For example, if we have 42 dimensional features, hence 42 dimensional mean vectors, we take the mean vector from the first component, the mean vector from the second component, and concatenate them to form a vector that has $42 + 42 = 84$ dimensions. We then concatenate the latter vector with the mean vector from the third component to form a vector that has $84 + 42 = 126$ dimensions. We continue until we have concatenated the mean vectors from all the components. If there are 1024 components, the supervector has 43,008 dimensions. We generate one supervector for each recording of each speaker.

Next, we reduce the number of dimensions using *principal component analysis* (PCA). PCA finds new dimensions which are linear combinations of the original dimensions such that the first PCA dimension accounts for the largest amount of variance in the training data, the second PCA dimension accounts for the largest remaining amount of variance in the training data after the first dimension is removed, the third PCA dimension accounts for the largest remaining amount of variance in the training data after the first and second dimensions are removed, etc. The number of dimensions is reduced by only using the first few PCA dimensions. These dimensions capture most of the variance in the data. Figure 6 shows an example of a reduction from two original dimensions to one PCA dimension. The PCA dimension is in the direction of maximum variance in the original two-dimensional space. (The initial development of PCA is credited to Pearson, 1901.)

---

[4] Enzinger (2016) ch. 4, obtained favorable results when a relatively small amount of case-specific data were used for training the UBM, i.e., data that represented the relevant population for the case and reflected the conditions of the questioned- and known-speaker recordings in the case. One should be cautious, however, because training with small amounts of data may give unstable results.

<Insert Figure 6 about here>

The supervectors are reduced from tens of thousands of dimensions to a much smaller number of dimensions, e.g., 400 dimensions. The reduced-dimension vectors are the i-vectors. The PCA dimensions are in the directions of maximum variance in the original space, irrespective of whether the variance is primarily due to speaker or condition (or other) differences, hence the result is sometimes called the *total variability space*.

Conceptually, i-vectors are the result of a mapping from supervectors onto a linear subspace with a lower number of dimensions. This is represented in Equation 9, in which $\boldsymbol{s}$ is a recording-specific supervector, $\boldsymbol{m}$ is the supervector for the UBM, and $\boldsymbol{v}$ is the i-vector for the specific recording. $\mathbf{T}$ is a low-rank matrix representing the linear subspace of the supervector space in which the i-vectors reside. The $\mathbf{T}$ matrix is designed such that the i-vectors have a few hundred dimensions rather than the tens of thousands of dimensions of the supervectors.

(9)
$$\boldsymbol{s} = \boldsymbol{m} + \mathbf{T}\boldsymbol{v}$$

The more commonly used procedures for extracting i-vectors do not use PCA for dimension reduction, but instead a form of factor analysis. The $\mathbf{T}$ matrix is trained using an iterative maximum likelihood technique (the EM algorithm). Rather than first adapting a GMM from the UBM, then concatenating a supervector, and then reducing its dimensions, a more direct procedure is used to map from feature vectors to i-vectors. The first step is to calculate the *Baum-Welch statistics* for each recording. The 0th order Baum-Welch statistics are based on the probability that a feature vector of a recording $j$ would belong to component $g$ of the UBM (like the responsibilities in the EM algorithm). For each feature vector, one value is calculated for each of the $G$ components, thus we have a matrix with $N_j$ columns (as many columns as there are feature vectors in the recording) and $G$ rows of responsibilities. We then sum over the columns to create a set of $G$ 0th order statistics for recording $j$. The 1st order Baum-Welch statistics are based on the deviations of the feature vectors of a recording from the mean vector of each component $g$ of the UBM, weighted by the probabilities that the feature vectors of the recording would belong to component $g$ (the responsibilities of component $g$ for the feature vectors). Deviations are calculated on a per-feature-dimension basis, hence for each of the $N_j$ feature vectors, $M$ values are calculated for each of the $G$ components (where $M$ is the number of feature dimensions). We thus have an $N_j$ by $M$ by $G$ matrix. We sum over the first of these dimensions to arrive at a set of 1st order statistics for recording $j$, consisting of $G$ vectors each of length $M$. The **T** matrix is trained using the 0th and 1st order Baum-Welch statistics from a set of training recordings, and an i-vector for a given recording can then be directly calculated using the **T** matrix and the 0th and 1st order Baum-Welch statistics from that recording. We provide mathematical details in Appendix A. For greater detail, see the references cited at the beginning of this section.

i-vectors are usually "whitened", i.e., subjected to radial Gaussianization and length normalization, so that they conform better to the assumptions of subsequent statistical modeling procedures (see García-Romero and Espy-Wilson, 2009). After whitening it is only the direction of an i-vector away from the origin of the i-vector space that is relevant (whitened i-vectors define points that lie on the hypersurface of a hypersphere).

## 5.2    i-vector domain mismatch compensation (LDA)

A common i-vector domain mismatch compensation technique is known in the automatic-speaker-recognition literature as *linear discriminant analysis* (LDA). In practice, linear discriminant analysis is not actually performed (i.e., neither likelihoods, nor posterior probabilities, not classification results are output), but *canonical linear discriminant functions* (CLDFs) are calculated and used to transform and reduce the dimensionality of the i-vectors (see Klecka, 1980; Fisher, 1936, is credited with the initial development of discriminant analysis and discriminant functions). In contrast to PCA which finds new dimensions that maximize the total variance, LDA finds dimensions that maximize the ratio of between-category variance versus within-category variance. CLDFs are trained, using i-vectors from training data that include multiple recordings of each speaker, with different recordings in different conditions. Each speaker is treated as a category, thus the CLDFs maximize the ratio of between-speaker variance to within-speaker variance, much of the within-speaker variance being due to variability in speaking styles and recording conditions. Only the first few CLDFs are used so as to primarily capture between-speaker variance. The CLDFs are used to transform the i-vectors into a new smaller set of dimensions, e.g., 50 dimensions rather than 400.

Figure 7 shows an example of CLDF reduction from two dimensions to one dimension. The same data are used for Figures 6 and 7 in order to illustrate the differences between the PCA and CLDF procedures. CLDFs could be calculated for supervectors rather than i-vectors, but there can be problems with attempting to train CLDFs in such a high-dimensional space.

<Insert Figure 7 about here>

### 5.3 PLDA

An i-vector approach produces a single i-vector for each recording. Hence, unlike in the case of GMM-UBM, there is a single vector for the known-speaker recording rather than a set of vectors that could be used to train a known-speaker model. A different modeling approach is therefore used to calculate likelihood ratios from i-vectors. In the automatic-speaker-recognition literature, this is known as *probabilistic linear discriminant analysis* (PLDA). PLDA is described in: Prince and Elder (2007); Kenny (2010); Brümmer and de Villiers (2010); Sizov et al. (2014).

PLDA is a *common-source model*,[5] i.e., it does not build a model for the specific known-speaker recording in the case, but instead answers the following two-part question:

> What would be the probability of obtaining the feature vectors of the questioned-speaker recording and the feature vectors of the known-speaker recording if the questioned speaker and the known speaker were the same speaker, a speaker selected at random from the relevant population?
>
> versus
>
> What would be the probability of obtaining the feature vectors of the questioned-speaker recording and the feature vectors of the known-speaker recording if the questioned speaker and

---

[5] For a discussion of the distinction between *specific-source* and *common-source* models, see Ommen and Saunders (2018).

the known speaker were different speakers, each selected at random from the relevant population?[6]

Conceptually, the answer to the first part of the question provides the numerator for the likelihood ratio and the answer to the second part provides the denominator.

Assume that we have a sample of $N$ speakers from the relevant population with multiple recordings of each speaker, and that some recordings of each sample speaker are in the questioned-speaker condition and others are in the known-speaker condition. We calculate an i-vector for each recording and apply i-vector domain mismatch compensation. Imagine that we also have a known-speaker recording and a questioned-speaker recording, and we calculate an i-vector for each of the two recordings and apply i-vector domain mismatch compensation, resulting in i-vectors $\boldsymbol{v}_k$ and $\boldsymbol{v}_q$ respectively.

In practice, we fit a multivariate model, but to simplify the explanation below we will describe a univariate model. We assume that both the within-speaker and between-speaker distributions are Gaussian. We also assume equal within-speaker variance for all speakers.

If $v_k$ and $v_q$ came from the same speaker ($q = k$), they would both come from a within-speaker distribution with a mean $\mu_k$ and variance $\sigma_k^2$. If they came from different speakers ($q \neq k$), one would come from a within-speaker distribution with a mean $\mu_k$ and variance $\sigma_k^2$, and the other from a within-speaker distribution with a mean $\mu_q$ and variance $\sigma_q^2$. We want to calculate a likelihood ratio as shown in Equations 10–12, i.e., the joint likelihood for $v_k$ and $v_q$ if they both came from the same within-speaker distribution versus the joint likelihood for $v_k$ and $v_q$ if each came from a different within-speaker distribution. If we assume equal variance for all speakers, including the questioned- and known-speaker, we can pool the i-vectors from all speakers in the sample of the population, calculate the pooled within-speaker variance, $\hat{\sigma}_w^2$, and then substitute that for $\sigma_k^2$ and $\sigma_q^2$, as in Equation 12.

---

[6] An objection may be raised that the known speaker was not selected at random, but on the basis of other evidence. The forensic practitioner's task, however, is to assess the strength of the particular evidence they have been asked to analyze – they should not consider the other evidence in the case, considering all the evidence is the role of the trier of fact. For the forensic practitioner's task, it is therefore acceptable to make the statistical assumption that the known speaker was selected at random from the relevant population.

(10)

$$\Lambda_{q,k} = \frac{f(v_q, v_k | H_{q=k})}{f(v_q, v_k | H_{q \neq k})}$$

(11)

$$\Lambda_{q,k} = \frac{f(v_q | \mu_k, \sigma_k^2) f(v_k | \mu_k, \sigma_k^2)}{f(v_q | \mu_q, \sigma_q^2) f(v_k | \mu_k, \sigma_k^2)}$$

(12)

$$\Lambda_{q,k} = \frac{f(v_q | \mu_k, \hat{\sigma}_w^2) f(v_k | \mu_k, \hat{\sigma}_w^2)}{f(v_q | \mu_q, \hat{\sigma}_w^2) f(v_k | \mu_k, \hat{\sigma}_w^2)}$$

We do not know the values for $\mu_k$ and $\mu_q$. For each speaker, $r$, in the sample of the relevant population, let us calculate the mean of the speaker's i-vectors: $\hat{\mu}_r$, $r \in \{1 \ldots N\}$. Since we do not know the value for $\mu_k$, let us randomly pick one of the speakers from the sample of the population and use their mean $\hat{\mu}_i$, and since we do not know the value for $\mu_q$, let us randomly pick another of the speakers from the population and use their mean $\hat{\mu}_j$. We impose the condition that $i \neq j$. We can now calculate a likelihood ratio as in Equation 13.

(13)

$$\Lambda_{q,k,i,j} = \frac{f(v_q | \hat{\mu}_i, \hat{\sigma}_w^2) f(v_k | \hat{\mu}_i, \hat{\sigma}_w^2)}{f(v_q | \hat{\mu}_j, \hat{\sigma}_w^2) f(v_k | \hat{\mu}_i, \hat{\sigma}_w^2)}$$

If we pick two more speakers at random, we will calculate a different value for the likelihood ratio. Let us imagine that we pick lots of random pairs of speakers and consider the average likelihood-ratio value.

If $v_k$ and $v_q$ are far apart, then the numerator of the likelihood ratio will always be relatively small because $\mu_i$ cannot be close to both $v_k$ and $v_q$. The denominator of the likelihood ratio will sometimes be relatively high because it is possible for $\mu_i$ to be close to $v_k$ and $\mu_j$ to be close to $v_q$. On average, the denominator will be larger than the numerator and the average value of the likelihood ratio will therefore be low.

forensic speech statistics - 2020-06-12a GSM                                                                                                           Page 28 of 65If $v_k$ and $v_q$ are close to each other and atypical with respect to the population, i.e., both out on the same tail of the distribution, then the value of the numerator of the likelihood ratio will usually be small because the probability of $\mu_i$ being close to $v_k$ and $v_q$ is small, but when $\mu_i$ is close to $v_k$ and $v_q$ the value of the numerator will be large. On average the value of the denominator will be even lower because the probability of both $\mu_i$ and $\mu_j$ being out on the same tail of the distribution and therefore one being close to $v_k$ and the other being close to $v_q$ will be lower than the probability of only $\mu_i$ being out on that tail of the distribution. On average, the numerator will be larger than the denominator and the average value of the likelihood ratio will therefore be high.

If $v_k$ and $v_q$ are close to each other but typical with respect to the population, i.e., both in the middle of the distribution, then the value of the numerator of the likelihood ratio will usually be large because the probability of $\mu_i$ being in the middle of the distribution and therefore being close to $v_k$ and $v_q$ will be large, but the value of the denominator will also usually be large because the probability of both $\mu_i$ and $\mu_j$ being in the middle of the distribution and therefore one being close to $v_k$ and the other being close to $v_q$ will also be large. On average, the values of numerator and denominator will be about the same and the average value of the likelihood ratio will therefore be close to 1.

In general, the average calculated value of the likelihood ratio will reflect how similar $v_k$ and $v_q$ are to each other, and how typical they are with respect to the sample of the population.

Rather than selecting pairs of speakers at random, we could systematically go through all possible combinations of speakers in our sample of the population and calculate the mean likelihood ratio, as in Equation 14 (the sum in parenthesis in the denominator is from 1 to $N-1$ since $j=i$ is skipped).

(14)

$$\Lambda_{q,k} = \frac{\frac{1}{N}\sum_{i=1}^{N}\left(f(v_q|\hat{\mu}_i, \hat{\sigma}_w^2)f(v_k|\hat{\mu}_i, \hat{\sigma}_w^2)\right)}{\frac{1}{N}\sum_{i=1}^{N}\left(f(v_k|\hat{\mu}_i, \hat{\sigma}_w^2)\frac{1}{N-1}\sum_{j=1}^{N-1}f(v_q|\hat{\mu}_j, \hat{\sigma}_w^2), j \neq i\right)}$$

Rather than a discrete summation using the mean of each speaker in the sample of the relevant population, we can model the distribution of the speaker means and use integration. The speaker means are treated as *nuisance parameters* and "integrated out". We calculate the mean and the variance of the speaker means: $\hat{\mu}_b$ and $\hat{\sigma}_b^2$ – this gives us the between-speaker distribution. Converting from discrete summation to integration, we arrive at Equation 15. For Gaussian distributions, there is a closed-form solution for the integrals, as given in Equation 16. Equation 16 uses bivariate Gaussian distributions. The covariance matrix in the numerator has positive off-diagonal elements, hence if the values of $v_k$ and $v_q$ are similar, e.g., both high or both low, the likelihood is greater. In contrast, the covariance matrix in the denominator is diagonal only, hence the likelihood is not greater if similarity is greater, e.g., the likelihood would be the same if both $v_k$ and $v_q$ are high or if one is high and the other is low. The smaller the within-speaker variance, $\hat{\sigma}_w^2$, the larger the off-diagonal elements, $\hat{\sigma}_b^2$, will be relative to the on-diagonal elements, $\hat{\sigma}_w^2 + \hat{\sigma}_b^2$, and the greater the effect of similarity on the value of the likelihood ratio.

(15)
$$\Lambda_{q,k} = \frac{\int f(v_q|\mu_r, \hat{\sigma}_w^2) f(v_k|\mu_r, \hat{\sigma}_w^2) f(\mu_r|\hat{\mu}_b, \hat{\sigma}_b^2) d\mu_r}{\int f(v_q|\mu_r, \hat{\sigma}_w^2) f(\mu_r|\hat{\mu}_b, \hat{\sigma}_b^2) d\mu_r \int f(v_k|\mu_r, \hat{\sigma}_w^2) f(\mu_r|\hat{\mu}_b, \hat{\sigma}_b^2) d\mu_r}$$

(16)
$$\Lambda_{q,k} = \frac{f\left(\begin{bmatrix}v_q\\v_k\end{bmatrix} \middle| \begin{bmatrix}\hat{\mu}_b\\\hat{\mu}_b\end{bmatrix}, \begin{bmatrix}\hat{\sigma}_w^2 + \hat{\sigma}_b^2 & \hat{\sigma}_b^2 \\ \hat{\sigma}_b^2 & \hat{\sigma}_w^2 + \hat{\sigma}_b^2\end{bmatrix}\right)}{f(v_q|\hat{\mu}_b, \hat{\sigma}_w^2 + \hat{\sigma}_b^2) f(v_k|\hat{\mu}_b, \hat{\sigma}_w^2 + \hat{\sigma}_b^2)} = \frac{f\left(\begin{bmatrix}v_q\\v_k\end{bmatrix} \middle| \begin{bmatrix}\hat{\mu}_b\\\hat{\mu}_b\end{bmatrix}, \begin{bmatrix}\hat{\sigma}_w^2 + \hat{\sigma}_b^2 & \hat{\sigma}_b^2 \\ \hat{\sigma}_b^2 & \hat{\sigma}_w^2 + \hat{\sigma}_b^2\end{bmatrix}\right)}{f\left(\begin{bmatrix}v_q\\v_k\end{bmatrix} \middle| \begin{bmatrix}\hat{\mu}_b\\\hat{\mu}_b\end{bmatrix}, \begin{bmatrix}\hat{\sigma}_w^2 + \hat{\sigma}_b^2 & 0 \\ 0 & \hat{\sigma}_w^2 + \hat{\sigma}_b^2\end{bmatrix}\right)}$$

Example PLDA numerator and denominator distributions are shown in Figure 8: values of $\hat{\mu}_b = 0$, $\hat{\sigma}_w^2 = 0.25$, $\hat{\sigma}_b^2 = 1$, $v_k = -1.5$, and $v_q = -1$ were selected for illustrative purposes only; the calculated likelihood-ratio value is 2.4.

<Insert Figure 8 about here>

To calculate the likelihood ratio for actual i-vectors, which are multivariate vectors, we use Equation 17, which is the multivariate equivalent of Equation 16 ($\widehat{\boldsymbol{\Sigma}}_w$ and $\widehat{\boldsymbol{\Sigma}}_b$ are within- and between-speaker covariance matrices respectively).

(17)

$$\Lambda_{q,k} = \frac{f\left(\begin{bmatrix}\boldsymbol{v}_q\\\boldsymbol{v}_k\end{bmatrix}\bigg|\begin{bmatrix}\widehat{\boldsymbol{\mu}}_b\\\widehat{\boldsymbol{\mu}}_b\end{bmatrix}, \begin{bmatrix}\widehat{\boldsymbol{\Sigma}}_w + \widehat{\boldsymbol{\Sigma}}_b & \widehat{\boldsymbol{\Sigma}}_b\\\widehat{\boldsymbol{\Sigma}}_b & \widehat{\boldsymbol{\Sigma}}_w + \widehat{\boldsymbol{\Sigma}}_b\end{bmatrix}\right)}{f(\boldsymbol{v}_q|\widehat{\boldsymbol{\mu}}_b, \widehat{\boldsymbol{\Sigma}}_w + \widehat{\boldsymbol{\Sigma}}_b)f(\boldsymbol{v}_k|\widehat{\boldsymbol{\mu}}_b, \widehat{\boldsymbol{\Sigma}}_w + \widehat{\boldsymbol{\Sigma}}_b)}$$

In practice, the output of PLDA is treated as a score and subjected to score-to-likelihood-ratio conversion, see Section 7.

What we have described above is known as the *two-covariance* version of PLDA. There are at least two other versions of PLDA (that Sizov et al., 2014, label *standard* and *simplified*). The latter are more complex than the two-covariance version in that they include dimension reduction to work in speaker-variability and session-variability subspaces. Since the latter include subspace modeling, they are not usually preceded by CLDF transformation, whereas since the two-covariance version does not include subspace modeling it should be preceded by CLDF transformation. For further details of the two-covariance version of PLDA and for details of the standard and simplified versions of PLDA, see the references cited at the beginning of this section. For a comparison of the three, see Sizov et al. (2014).

## 6     DNN-based systems

*Artificial neural networks* are machine learning models that consist of *nodes* (neurons) that are organized in *layers*, and *connections* between the nodes (synapses). An example of the architecture of an artificial neural network is shown in Figure 9. This is an example of a standard feedforward fully-connected architecture, i.e., all units in a given layer are connected to all the units in the preceding layer and all the units in the following layer. The example has an input layer, two hidden layers, and an output layer. A hidden layer is any layer other than the input or output layer. The number of hidden layers is a design choice (this example happens to have two hidden layers). An artificial neural network with more than one hidden layer is known as a *deep neural network* (DNN). A value is presented at each input node, e.g., the set of input nodes accepts a feature vector, $\boldsymbol{x}$. Each input node then has an activation level, $h_{0,n_a}$, corresponding to an $x$ value. Each input node is connected to each node in the first hidden layer via connections each of which is given a weight, $w_{l_{j-1},n_a,l_j,n_b}$, where $l_{j-1}$ and $l_j$ index the layers (in this case the input layer, $l_{j-1} = 0$, and the first hidden layer, $l_j = 1$), and $n_a$ and $n_b$ index the particular nodes that are connected. Each node in the first hidden layer then has an activation level, $h_{l_j,n_b}$, that is a function, $\varphi$, of the weighted sum of the values presented to the input layer, see Equation 18 ($b_{l_j,n_b}$ is a node-specific bias term). A non-linear function, e.g., a sigmoidal function, is usually used.

(18)

$$h_{l_j,n_b} = \varphi \left( b_{l_j,n_b} + \sum_{n_a=1}^{N_a} w_{l_{j-1},n_a,l_j,n_b} h_{l_{j-1},n_a} \right)$$

The activation levels of nodes in higher layers are in turn based on weighted sums of the activation levels of the nodes in the immediately preceding layer.

<Insert Figure 9 about here>

Usually, the task of the network is to predict the category that the input belongs to, and each node in the output layer represents a category. For example, for an optical character recognition system, the input could be an image of a character (i.e., a letter of the alphabet, a number, or a punctuation mark), each output node would represent a particular character, and the relative level of activation of an output node would represent the strength of the model's prediction that the input image is of the character corresponding to that output node. All else being equal, in order to classify the image, one would select the character corresponding to the output node with the highest level of activation. The activations of the output nodes can be scaled to represent posterior probabilities.

Training an artificial neural network is the process of setting the connection weights (and node-specific biases) so as to optimize the network's performance on the classification task. A standard supervised training algorithm is *backpropagation*. This is an iterative process similar to the EM algorithm. Starting from an initial state (e.g., a random seeding of the weights), it involves presenting inputs for which the categories are known, comparing the activation levels of the output nodes with the desired activation levels given the known category of the input, and adjusting the weights so as to reduce the error, i.e., to lead to higher relative activation of the output node corresponding to the known category of the input. For more detailed introductions to artificial neural networks, including network training, see Duda et al. (2000) ch. 6, and Hastie et al. (2009) ch. 11.

Below, we briefly describe three DNN approaches: (1) *DNN senone posterior i-vector systems*, (2)

*Bottleneck-feature based systems*, and (3) *DNN speaker embedding (x-vector) systems*. All three were common in state-of-the-art automatic-speaker-recognition systems when we began writing the present chapter in 2017. By 2019 the x-vector approach had emerged as the clear winner.

## 6.1    DNN senone posterior i-vector systems

A senone posterior i-vector system can be considered a variant of an i-vector PLDA system, but using a DNN to train the UBM rather than using the EM algorithm (see García-Romero et al., 2014; Kenny et al., 2014; Lei et al., 2014). In contrast to standard i-vectors, senone posterior i-vectors are designed to explicitly capture information about how speakers pronounce speech sounds.

The input to the DNN consists of a series of feature vectors (e.g., MFCC + delta + double delta). The DNN is trained to classify the input into sequences of speech sounds, e.g., triphones. For simplicity, we present an example based on orthography rather than speech sounds: The word "forensic" contains the trigraphs "for", "ore", "ren", "ens", "nsi", and "sic". In general, the sequences of speech sounds are called *senones*. (Although there has been a shift and broadening of meaning to its current use in automatic speaker recognition, the term "senone" was originally coined in Hwang and Huang, 1992, to mean a subphonetic unit.)

Standard supervised training techniques are used to train the DNN to classify a large number of different senones (e.g., between 5000 and 10,000). A "true" category label for the senone corresponding to each feature vector is typically computed using an automatic-speech-recognition system. The DNN is trained using a large number of recordings for a large number of speakers in diverse recording conditions. The activation of each output node of the DNN is scaled to represent the posterior probability for the senone it was trained on.

The UBM consist of a GMM in which each component Gaussian corresponds to a DNN output node. Feature vectors from a new set of multiple recordings of multiple speakers are presented to the DNN. The activation of each DNN output node, $g$, is obtained for each feature vector, $x_i$, and is used as the responsibility, $\gamma_{g,i}$, in Equations 3–5 (the version of Equations 3–5 given in Section 4.1 was for univariate data, the corresponding multivariate version is actually used). The $x_i$ and $\gamma_{g,i}$ provide all the information necessary to train the UBM in a single iteration.

## 6.2 Bottleneck-feature based systems

In a bottleneck-feature based system (see Yaman et al., 2012; García-Romero and McCree, 2015; Lozano-Díez et al., 2016; Matějka et al., 2016) a DNN is trained in the same way as in the senone posterior i-vector system but one layer in the DNN, the *bottleneck layer*, has a substantially smaller number of nodes than the other hidden layers, e.g., 60–80 nodes in the bottleneck layer compared to 1000–1500 nodes in other layers. The bottleneck layer is designed to capture information about the phonetic content of voice recordings. The small number of nodes provides a compact representation of this information. The bottleneck layer may be between other hidden layers, or may be immediately before the output layer.

For each input feature vector (e.g., MFCC + delta + double delta), the activations of the nodes in the bottleneck layer are used as a new feature vector, a *bottleneck-feature vector*. The bottleneck-feature vectors are usually concatenated with the original MFCC + delta + double delta feature vectors and the concatenated feature vectors then used as input to a standard UBM-based i-vector PLDA system (note that the activations of the DNN's output layer are not used).

## 6.3 DNN speaker embedding systems (x-vector systems)

A DNN speaker embedding system is described in Snyder et al. (2017), and specific values given below are those of that system (other variants are described in: Peddinti, Chen, et al., 2015; Peddinti, Povey, and Khudanpur, 2015; Snyder et al., 2018; Lee et al., 2020; Matějka et al., 2020; Villalba et al., 2020). In this context, *embeddings* are fixed-length vectors that are used instead of i-vectors as input to PLDA. These embeddings are also known as *x-vectors*. In contrast to senone posterior i-vector systems and bottleneck-feature systems that are trained to classify senones, a DNN embedding system is trained to classify speakers.

Rather than the input layer of a DNN speaker embedding system accepting a single feature vector, it is organized to accept a two-dimensional matrix in which one dimension is the elements of the MFCC vectors (that represent the frequency components of the signal at the time corresponding to each vector), and the other dimension is time. Time is discretized according to the frame shifts used when extracting the series of MFCC vectors from the voice recording. Deltas and double deltas are not included in the input feature vectors, instead each node in the input layer is connected to the MFCC vectors from 5 contiguous frames, i.e., the MFCC vectors corresponding to the frames at times $t$–2, $t$–1, $t$, $t$+1, $t$+2. Nodes in the first hidden layer are connected to nodes at times $t$–2, $t$, $t$+2 of the input layer, and nodes in the second hidden layer are connected to nodes at times $t$–3, $t$, $t$+3 of the first hidden layer, see Figure 10. The input layer ultimately spans frames $t$–7 to $t$+7.[7] The second hidden layer is followed by two more hidden layers that are unidimensional – their nodes are only connected to nodes at time $t$ of the preceding layer, hence the time dimension has been collapsed. All the layers described so far are called *frame-level layers*. Layers that receive inputs from multiple times are said to *splice* those inputs together.

<Insert Figure 10 about here>

The next layer in the DNN is called a *statistics-pooling layer*. The whole sequence of MFCC vectors from the speech of the speaker of interest in a recording is presented to the system, the input layer sees 15 frames at a time (frames $t$–7 through $t$+7) and $t$ is advanced one frame at a time through the entire sequence of MFCC vectors. For each node in the immediately preceding layer, the statistics-pooling layer has two nodes, one calculates the mean of the activation of the preceding layer's node over all $t$, and the other calculates the standard deviation of the activation of the preceding layer's node over all $t$. The statistics-pooling layer provides a fixed-length vector irrespective of the length of the recording.

The values of the statistics-pooling layer are used as an input vector to a set of three more layers, including the output layer. These are called *segment-level layers*.[8] The second of these layers has fewer nodes than the first, e.g., 300 versus 512 nodes, hence the second is a bottleneck. The activations of either the first layer or the second layer (or both) can be used as an embedding (an x-vector). As with i-vectors, CLDF for mismatch compensation and dimension reduction can be applied before PLDA.

Training the DNN makes use of recordings of several thousand speakers. Multiple recordings of each speaker are used, with different recordings reflecting different conditions. In addition to recordings genuinely recorded under different conditions, the range of variability is often increased by entering the same recordings multiple times after they have undergone processing to simulate multiple other conditions, e.g., addition of different types and levels of noise, lossy compression, and reverberation (see McLaren et al., 2018). The aim is for the DNN to learn the speaker-dependent properties of the input and to learn to ignore the condition-dependent properties. The activations of the nodes in the output layer indicate posterior probabilities for the speakers in the training set. The ultimate aim, however, is not to build a system that classifies those particular speakers, hence the activations of the DNN's output layer are not used as input to the PLDA. Instead, the aim is to generate fixed-length vectors that characterize the different speech properties of different speakers in general. Those are the embeddings (x-vectors), and it is the embeddings (x-vectors) that are used as input to the PLDA.

---

[7] The weights for the connections from the input layer to the nodes in the first hidden layer are not independent of each other. The weights on different connections entering a first-layer node generally differ from each other, but the same set of weights are used for the set of connections entering each and every first-layer node.

[8] In the automatic-speaker-recognition literature, the term "segment" refers to all the speech of the speaker of interest in the whole recording. It is the latter meaning which is intended, as opposed to the meaning of "segment" in phonetics, which is an individual speech sound, a realization of a phoneme.

## 7    Score-to-likelihood-ratio conversion (calibration)

*Score-to-likelihood-ratio conversion* (aka *calibration*), particularly using logistic regression, is described in: Pigeon et al. (2000), Ramos Castro (2007), González-Rodríguez et al. (2007), Morrison (2013), Morrison and Poh (2018), Ferrer et al. (2019).

Scores are similar to likelihood ratios in that, for a specific-source model, they take account of the similarity of the voice on the questioned-speaker recording with respect to the voice of the known speaker and the typicality of the voice on the questioned-speaker recording with respect to the voices of speakers in the relevant population, or, for a common-source model, they take account of the similarity of the voices on the questioned- and known-speaker recordings and their typicality with respect to the voices of speakers in the relevant population.[9]

In forensic voice comparison, scores that take account of both similarity and typicality are generated using models such as GMM-UBM and PLDA. Because of violations of modeling assumptions or lack of sufficient training data, however, scores are not interpretable as meaningful likelihood-ratio values answering the question posed by the same- and different-speaker hypotheses in the case. An additional step is needed to convert scores to interpretable likelihood ratios. Another way of looking at the problem is that scores are likelihood ratios, but their values are not interpretable because they are not calibrated. The score-to-likelihood-ratio-conversion process is therefore also known as calibration, and for brevity we will henceforth tend to use the latter term.

---

[9] Scores that are based only on similarity do not account for typicality with respect to the relevant population and this cannot be remedied at the score-to-likelihood-ratio-conversion stage (see Morrison and Enzinger, 2018).

In order to obtain data for training a calibration model, feature vectors from a large number of same-speaker pairs of recordings and a large number of different-speaker pairs of recordings are input to the GMM-UBM system, the i-vector PLDA system, or the DNN-based system, and a set of *same-speaker scores* and a set of *different-speaker scores* are obtained. The training data should come from a sample of the relevant population, and should be distinct from the data that were used to train earlier models in the system. Also, one member of each pair should have conditions reflecting those of the known-speaker recording and the other should have conditions reflecting those of the questioned-speaker recording. Training using data that do not reflect the relevant population and the conditions of the case will lead to miscalibrated results and poorer performance. Mandasari et al. (2013) and Mandasari et al. (2015) illustrated that training using recordings that matched the durations of the test recordings resulted in better performance than using longer-duration training recordings, and that training using recordings whose signal to noise ratios matched those of the test recordings resulted in better performance than using training recordings whose signal to noise ratios did not match those of the test recordings.[10]

A simple score-to-likelihood-ratio conversion model makes use of two Gaussian distributions with equal variance. We calculate the mean for the *same-speaker model*, $\hat{\mu}_s$, using the same-speaker training scores, and the mean for the *different-speaker model*, $\hat{\mu}_d$, using the different-speaker training scores. We calculate a single variance, $\hat{\sigma}^2$, using data pooled from both categories, i.e., a *pooled variance*. To calculate a likelihood ratio, we first calculate a score, $S_{q,k}$, for the comparison of the voices on the questioned-speaker and known-speaker recordings. We then calculate the likelihood of the same-speaker model at the value of $S_{q,k}$, the likelihood of the different-speaker model at the value of $S_{q,k}$, and divide the former by the latter, see Equation 19.

---

[10] The aim of Mandasari et al. (2013) and Mandasari et al. (2015) was to find automatic procedures to address this problem rather than relying on a human expert to select data on a case by case basis; see also McLaren et al. (2014).

(19)

$$\Lambda_{q,k} = \frac{f(S_{q,k}|\hat{\mu}_s, \hat{\sigma}^2)}{f(S_{q,k}|\hat{\mu}_d, \hat{\sigma}^2)}$$

The calculation of a likelihood ratio using this pooled-variance two-Gaussian model is illustrated in the top panel of Figure 11 in which, for illustrative purposes, $\hat{\mu}_s = +0.5$, $\hat{\mu}_d = -1.5$, and $\hat{\sigma}^2 = 1$. The likelihood-ratio value calculated for an illustrative test score of +0.5 is 0.399/0.054 = 7.39.

<Insert Figure 11 about here>

Note that whereas in the feature domain the data were multivariate and had complex distributions and we fitted models requiring a large number of parameter values to be estimated, in the score domain the data are univariate and we fit models requiring only a small number of parameter values to be estimated (e.g., two means and one variance). We can therefore obtain better parameter estimates in the score domain than in the feature domain, and hence the output of the score-to-likelihood-ratio conversion model is well calibrated.

The value of a score has no meaning by itself, but if one score has a higher value than another score then the higher-valued score indicates greater relative support for the same-speaker hypothesis over the different-speaker hypothesis than does the lower-valued score. Thus, although we do not know whether both scores correspond to likelihood ratios less than 1, or both to likelihood ratios greater than 1, or one to a likelihood ratio less than 1 and the other to a likelihood ratio greater than 1, and we do not know whether they correspond to likelihood ratios that are close to each other or far from each other, we do know that the higher-valued score corresponds to a higher-valued likelihood ratio than does the lower-valued score. The model for converting from scores to likelihood ratios should therefore be *monotonic*, that is, it should preserve the ranking of scores, a lower-ranked score should not be converted to a higher likelihood-ratio value than the likelihood-ratio value corresponding to a higher-ranked score. In the model described above, using two Gaussians with the same variance results in a monotonic conversion. Using a different variance for each Gaussian would not be monotonic.

The pooled-variance two-Gaussian model is a *linear discriminant analysis* model, and can be shown to reduce to a linear model in a logged-odds space or in a log-likelihood-ratio space. A linear model has the form $y = a + bx$, and requires the estimation of only two parameter values, an intercept, $a$, and a slope, $b$. If natural logarithms are used in calculating the score, the intercept and slope are as shown in Equations 20–22, and the output is the natural logarithm of the likelihood ratio.

(20)
$$\log(\Lambda_{q,k}) = a + bS_{q,k}$$

(21)
$$a = -\frac{\hat{\mu}_s^2 - \hat{\mu}_d^2}{2\hat{\sigma}^2} = -b(\hat{\mu}_s + \hat{\mu}_d)/2$$

(22)
$$b = \frac{\hat{\mu}_s - \hat{\mu}_d}{\hat{\sigma}^2}$$

This version of the model is illustrated in the middle panel of Figure 11, in which $a = +1$ and $b = +2$. The illustrative test score of +0.5 converts to a natural log likelihood ratio of $1 + 2 \times 0.5 = 2$, which as before is a likelihood ratio of 7.39.

In practice, it is more common to use *logistic regression* rather than linear discriminant analysis. If the assumptions of Gaussian distributions with equal variances hold, then logistic regression and linear discriminant analysis will give the same results (Hastie et al., 2009, §4.4.5). Logistic regression, however, is more robust to violations of those assumptions, thus in general it results in better performance. The same-speaker and the different-speaker scores are used to train the logistic-regression model (see Figure 11 bottom panel), which is linear in the log-likelihood-ratio space (see Figure 11 middle panel), hence training results in calculation of the values for the intercept, $a$, and slope, $b$. These values are then used in Equation 20 to convert the score value to a log likelihood-ratio value.

Logistic regression is trained using an iterative procedure that minimizes the deviance statistic (see Menard, 2010; Hosmer et al., 2013). It is usually used to calculate a posterior probability (see Figure 11 bottom panel), but by converting from posterior probability to posterior odds and training the model using equal priors for the same-speaker and the different-speaker categories, the output is interpretable as a log likelihood ratio (see Figure 11 middle panel). Logistic regression is a discriminative procedure rather than a generative procedure and hence does not provide separate values that could be interpreted as the numerator and denominator of the likelihood ratio. Its output is not literally the ratio of two likelihoods, but the interpretation of its output as a likelihood ratio is justified by its analogy with linear discriminant analysis.

# 8      Validation

Empirical validation of a forensic analysis system under conditions reflecting those of the case to which it is to be applied is required to inform admissibility decisions under United States Federal Rules of Evidence 702 and the *Daubert* trilogy of Supreme Court rulings (*Daubert v. Merrell Dow Pharmaceuticals*, 1993; *General Electric v. Joiner*, 1997; and *Kumho Tire v. Carmichael*, 1999), and under England and Wales Criminal Practice Directions (2015) section 19A. This has been emphasized by President Obama's Council of Advisors on Science and Technology (2016) and by the England and Wales Forensic Science Regulator (2014). As mentioned in the introduction, there have been calls from the 1960s onward for the performance of forensic-voice-comparison systems to be empirically validated under casework conditions (for a review see Morrison, 2014), and such validation is advocated by the ENFSI guidelines (Drygajlo et al., 2015).

The results of a validation study depend on both the system that is tested and the conditions under which it is tested. In order to provide information that will be helpful in understanding how well a forensic-voice-comparison system will perform when applied in a particular case, what must be tested is the system that will actually be used in the case, and it must be tested using test recordings that are sufficiently representative of the relevant population and sufficiently reflective of the conditions of the questioned- and known-speaker recordings in the case. Results of tests of the system with other populations or under other conditions (or tests of other systems) will in general not be informative as to the performance of the system when applied in the particular case, and could be highly misleading. Since there can be substantial variation in relevant population and recording conditions from case to case, case-specific validation may be required.

In black-box validation studies, pairs of recordings are entered into the forensic-voice-comparison system, and in response to each input pair the system outputs a likelihood ratio. In each pair, one member of the pair must have conditions that reflect those of the questioned-speaker recording and the other must have conditions that reflect those of the known-speaker recording. Some pairs must be same-speaker pairs, and others must be different-speaker pairs. For each pair, the goodness (or badness) of the likelihood-ratio output is assessed with respect to the tester's knowledge of whether the input was a same-speaker or a different-speaker pair. Traditional performance metrics such as false-alarm rate and miss rate depend on making binary decision by applying a threshold to a posterior probability, and are therefore inconsistent with the likelihood-ratio framework. An appropriate metric would be based directly on the likelihood-ratio value of the output and would take account of the magnitude of the likelihood-ratio value. Given a same-speaker input, a good output would be a likelihood-ratio value that is much larger than 1, a less good output would be a value that is only a little larger than 1, a bad output would be a value less than 1, and a worse output would be a value much less than 1. *Mutatis mutandis* for a different-speaker input for which a good output would be a value much less than 1.

In the forensic-voice-comparison literature, the log-likelihood-ratio cost ($C_{llr}$) is a popular metric for quantifying system performance (Brümmer and du Preez, 2006, González-Rodríguez et al., 2007, Morrison, 2011, Drygajlo et al., 2015, Morrison and Enzinger, 2016, Meuwly et al., 2017). The formula for calculating $C_{llr}$ is given in Equation 23, in which $\Lambda_s$ and $\Lambda_d$ are likelihood-ratio outputs corresponding to same- and different-speaker input pairs respectively, and $N_s$ and $N_d$ are the number of same- and different-speaker input pairs respectively.

(23)
$$C_{llr} = \frac{1}{2}\left(\frac{1}{N_s}\sum_{i=1}^{N_s}\log_2\left(1 + \frac{1}{\Lambda_{s_i}}\right) + \frac{1}{N_d}\sum_{j=1}^{N_d}\log_2\left(1 + \Lambda_{d_j}\right)\right)$$

Figure 12 plots the penalty functions for log likelihood-ratio outputs corresponding to same- and different-speaker input pairs. These are the functions within Equation 20's left and right summations respectively. If the input is a log likelihood-ratio from a same-speaker pair and its value is much greater than 0 it receives a small penalty value, but if its value is lower it receives a higher penalty value. If the input is a log likelihood-ratio from a different-speaker pair and its value is much less than 0 it receives a small penalty value, but if its value is higher it receives a higher penalty value. $C_{llr}$ is calculated as the mean of the penalty values with equal weight given to the set of same-speaker penalty values as to the set of different-speaker penalty values.

<Insert Figure 12 about here>

Smaller $C_{\text{llr}}$ values indicate better performance. $C_{\text{llr}}$ values generally lie in the range 0 to 1. $C_{\text{llr}}$ values cannot be less than or equal to 0. A system that gave no useful information and always responded with a likelihood ratio of 1, irrespective of the input, would have a $C_{\text{llr}}$ value of 1. $C_{\text{llr}}$ values substantially greater than 1 can be produced by uncalibrated or miscalibrated systems.

In the forensic-voice-comparison literature, a popular graphical representation of system performance is a Tippett plot (Meuwly, 2001; González-Rodríguez et al., 2007; Drygajlo et al., 2015; Morrison and Enzinger, 2016; Meuwly et al., 2017; Morrison and Enzinger, 2019). Tippett plots consist of plots of the empirical cumulative probability distribution of the likelihood-ratio values resulting from same-speaker inputs and of the empirical cumulative probability distribution of the likelihood-ratio values resulting from different-speaker inputs. The tradition is to plot lines joining the data points rather than to plot the data points themselves. Examples are shown in Figure 13. The *y*-axis values corresponding to the curves rising to the right give the proportion of same-speaker test results with log likelihood-ratio values less than or equal to the corresponding value on the *x*-axis. The *y*-axis values corresponding to the curves rising to the left give the proportion of different-speaker test results with log likelihood-ratio values greater than or equal to the corresponding value on the *x*-axis. In general, shallower curves with greater separation between the two curves indicates better performance. Tippett plots can also reveal problems such as bias in the output.

<Insert Figure 13 about here>

In the top panel of Figure 13 the separation between the same-speaker and different-speaker curves is small and the system is clearly biased – both the same-speaker and different-speaker curves are too far to the right. The middle panel of Figure 13 shows the results of a validation of a better performing system. This Tippett plot has somewhat greater separation between the same-speaker and different-speaker curves and the results are not obviously biased. The second system was actually the same as the first system except that it included a logistic-regression calibration stage whereas the first did not. The bottom panel of Figure 13 shows the results of a validation of a system with substantially better performance – the same-speaker and different-speaker curves have greater separation and are shallower. The $C_{\text{llr}}$ values calculated on the same validation results as shown in the top, middle, and bottom panels of Figure 13 are 1.068, 0.698, and 0.307 respectively.

Given the range of possible forensic-voice-comparison systems and the case-to-case variability in relevant population and in recording conditions, there is no validation study that is representative across systems and across casework conditions. Below we list a number of published studies that report on empirical validation of several different systems under several different sets of conditions (including a series of papers in a journal virtual special issue in which multiple different systems were tested under the same conditions). Full references are provided in the list. These references are not repeated in the main reference list at the end of the chapter unless the publications are referenced elsewhere in the chapter. The references are listed by year of publication, and those published in the same year are listed in alphabetical order. Publications are only included in the list if they report on empirical validation of human-supervised automatic forensic-voice-comparison systems under conditions that reasonably closely reflect forensic casework conditions, the systems output likelihood-ratio values, and the validation results are reported using metrics and graphics that are consistent with the likelihood-ratio framework. We apologize if we have inadvertently omitted any studies that meet these criteria. The list was current as of June 1, 2020. We expect more validation studies to be published in the future.

Unless the relevant population and conditions of a published study reflect those of a particular case and the system validated in the study is the one that will be used in that particular case, the published study should not be used in a debate as to whether testimony derived from that forensic-voice-comparison system should be admitted in that particular case. If no validation study exists that sufficiently closely reflects the relevant population and conditions for the case, a case-specific validation study will have to be conducted. We would also suggest that whether the level of performance of a system is sufficient should be decided on a case by case basis.[11] Poorer recording conditions will lead to poorer performance, but a system with relatively poor performance may still be capable of providing useful information to the court in the form of a moderately sized likelihood-ratio value – Tippett plots will give an indication of the range of possible likelihood-ratio values that the system could generate under the test conditions.

---

[11] At a consensus-development meeting held in September 2019, a number of experts in validation of forensic-voice-comparison systems concluded that the only logically justified validation threshold for $C_{llr}$ was 1. As of June 1, 2020, a written statement of the consensus is still under development.

## 8.1 List of published validation studies

## 9      Conclusion

We have described signal-processing and statistical-modeling techniques that are commonly used to calculate likelihood ratios in human-supervised automatic approaches to forensic voice comparison. We hope that we have fulfilled our aim of bridging the gap between general introductions to forensic voice comparison and the highly technical and often fragmented automatic-speaker-recognition literature from which the signal-processing and statistical-modeling techniques are mostly drawn. We hope that this has been of value to students of forensic voice comparison and to researchers interested in understanding statistical modeling techniques that could potentially also be applied to data from other branches of forensic science.

## 10     Acknowledgments

We dedicate this chapter in memory of David Lucy, who died on June 20, 2018. David was perhaps best known in the forensic science community for his book *Introduction to Statistics for Forensic Scientists* (Lucy, 2005) and for his work on statistical models for the calculation of likelihood ratios, including his work on multivariate kernel density models, e.g., Aitken and Lucy (2004). These models were used extensively by the first two authors of the current chapter in their earlier work on acoustic-phonetic approaches to forensic voice comparison. David will be greatly missed. We extend our condolences to his family and friends.

Earlier collaboration between the first four authors that ultimately led to the writing of this chapter benefited from support provided by the Australian Research Council, Australian Federal Police, New South Wales Police, Queensland Police, National Institute of Forensic Science, Australasian Speech Science and Technology Association, and the Guardia Civil through Linkage Project LP100200142. The contributions of the last three authors were supported by the Spanish Ministry of Economy and Competitiveness through project TEC2015-68172-C2-1-P, and by collaboration over the last 20 years between AUDIAS and the Guardia Civil. Contributions by the first two authors to late stages of the preparation of the chapter were supported by Research England's Expanding Excellence in England Fund as part of funding for the Aston Institute for Forensic Linguistics 2019–2022.

# 11 Appendix A: Mathematical details of T matrix training and i-vector extraction

This appendix provides details of the procedures usually used for $\mathbf{T}$ matrix training and extraction of i-vectors. It is based on the factor analysis procedure described in Dehak et al. (2011).

1. Feature vectors from all the speakers in the training set are pooled and then used to train a UBM with the parameter set $\Omega = \{w_g, \boldsymbol{\mu}_g, \boldsymbol{\Sigma}_g\}_{g=1}^{G}$, i.e., $G$ Gaussians with mean vectors $\boldsymbol{\mu}_g$, diagonal-only covariance matrices $\boldsymbol{\Sigma}_g$, and weights $w_g$.

2. For each recording $j$ in the training and test sets, centralized 0th and 1st order Baum-Welch statistics, $n_{g,j}$ and $\mathbf{f}_{g,j}$, are computed with respect to the UBM, see Equations A1–A3 in which: $x_{j_i}$ is the $i$th feature vector of recording $j$; $\gamma_{g,i}$ is a responsibility (Equation A3 is the multivariate equivalent of Equation 2 in Section 4.1); $N_j$ is the number of feature vectors extracted from recording $j$; $n_{g,j}$ is the probability that the feature vectors of recording $j$ would belong to Gaussian $g$ of the UBM; and $\mathbf{f}_{g,j}$ is the sum, over all vectors in the recording, of the deviation of each vector from the UBM mean vector, weighted by the probability that it belongs to Gaussian $g$.

(A1)
$$n_{g,j} = \sum_{i=1}^{N_j} \gamma_{g,i}$$

(A2)
$$\mathbf{f}_{g,j} = \sum_{i=1}^{N_j} \gamma_{g,i} \left(x_{j_i} - \boldsymbol{\mu}_g\right)$$

(A3)
$$\gamma_{g,i} = \frac{w_g f\left(x_{j_i} | \boldsymbol{\mu}_g, \boldsymbol{\Sigma}_g\right)}{\sum_{l=1}^{G} w_l f\left(x_{j_i} | \boldsymbol{\mu}_l, \boldsymbol{\Sigma}_l\right)}$$

3. A maximum-likelihood estimate of the $\mathbf{T}$ matrix is trained using the training data and an EM algorithm (Kenny et al., 2005; Matrouf et al., 2007). First, each element of the $\mathbf{T}$ matrix is initialized by random draws from the standard Gaussian distribution. In the expectation step, for each recording $j \in \{1 \ldots J\}$ in the training set, we evaluate the posterior distribution of $\phi_j$ given the Baum-Welch statistics $\{n_{g,j}, \mathbf{f}_{g,j}\}_{g=1}^{G}$ from that recording. Assuming a standard Gaussian prior for $\phi_j$ and using the current estimate of the $\mathbf{T}$ matrix (denoted as $\mathbf{T}_{\text{old}}$), the first and second moments, $\langle \phi_j \rangle$ and $\langle \phi_j \phi_j^t \rangle$, of the posterior are calculated as in Equations A4 and A5. $\mathbf{L}_j^{-1}$ is the posterior covariance matrix. $\mathbf{\Sigma}_g$ is the covariance matrix of the $g$th Gaussian of the UBM. In the maximization step, the $\mathbf{T}$ matrix is updated as in Equation A7 (see Proposition 3 in Kenny et al., 2005, §III). Often, a minimum divergence step is added after the maximization step, in which the $\mathbf{T}$ matrix is updated so that the empirical i-vector distribution better conforms to a standard Gaussian prior (Kenny, 2008, §II-B; Glembek, 2012, §3.6.4). This is achieved by computing the dot-product of the matrix $\mathbf{T}_{g,\text{ML}}$ after the maximization step and the matrix $\mathbf{Q}$ obtained using a symmetrical decomposition (e.g., Cholesky decomposition) of $\mathbf{P}^{-1}$, the covariance matrix of the prior, Equations A8–A10. Empirically, adding the minimum divergence step results in faster convergence of the training algorithm. The expectation-maximization and minimum divergence steps are repeated for multiple iterations.

(A4)
$$\mathbf{L}_j = \mathbf{I} + \sum_{g=1}^{G} n_{g,j} \mathbf{T}_{g,\text{old}}^t \mathbf{\Sigma}_g^{-1} \mathbf{T}_{g,\text{old}}$$

(A5)
$$\langle \phi_j \rangle = \mathbf{L}_j^{-1} \sum_{g=1}^{G} \mathbf{T}_{g,\text{old}}^t \mathbf{\Sigma}_g^{-1} \mathbf{f}_{g,j}$$

(A6)
$$\langle \phi_j \phi_j^t \rangle = \mathbf{L}_j^{-1} + \langle \phi_j \rangle \langle \phi_j \rangle^t$$

(A7)
$$\mathbf{T}_{g,\text{ML}} = \left( \sum_{j=1}^{J} n_{g,j} \langle \phi_j \phi_j^t \rangle \right)^{-1} \left( \sum_{j=1}^{J} \mathbf{f}_{g,j} \langle \phi_j \rangle^t \right)$$

(A8)

$$\mathbf{P}^{-1} = \frac{1}{J} \sum_{j=1}^{J} \mathbf{L}_j^{-1} + \langle \phi_j \rangle \langle \phi_j \rangle^t$$

(A9)

$$\mathbf{Q}\mathbf{Q}^t = \mathbf{P}^{-1}$$

(A10)

$$\mathbf{T}_{g,\text{MD}} = \mathbf{T}_{g,\text{ML}} \, \mathbf{Q}$$

(A11)

$$\mathbf{T}_{\text{new}} = \begin{bmatrix} \mathbf{T}_{1,\text{MD}} \\ \vdots \\ \mathbf{T}_{G,\text{MD}} \end{bmatrix}$$

4. For each recording, $j$, in the training and test sets, an i-vector is obtained as the posterior mean $\langle \phi_j \rangle$ using the Baum-Welch statistics $\{n_{g,j}, \mathbf{f}_{g,j}\}_{g=1}^{G}$ computed from recording $j$. The calculations are the same as in Equations A4 and A5, but using the final estimate of the $\mathbf{T}$ matrix.

forensic speech statistics - 2020-06-12a GSM                                                                                      Page 59 of 65Pearson, K. 1901. On lines and planes of closest fit to systems of points in space. *The London, Edinburgh, and Dublin Philosophical Magazine and Journal of Science, Series 6*, 2(11): 559–572. http://dx.doi.org/10.1080/14786440109462720

Pigeon, S., Druyts, P., and Verlinde, P. 2000. Applying logistic regression to the fusion of the NIST'99 1-speaker submissions. *Digital Signal Processing*, 10: 237–248. http://dx.doi.org/10.1006/dspr.1999.0358

President's Council of Advisors on Science and Technology 2016. *Forensic Science in Criminal Courts: Ensuring Scientific Validity of Feature-Comparison Methods*. https://obamawhitehouse.archives.gov/administration/eop/ostp/pcast/docsreports/

Prince, S.J.D. and Elder, J.H. 2007. Probabilistic linear discriminant analysis for inferences about identity. In *Proceedings of the IEEE 11th International Conference on Computer Vision*, pp. 1–8. https://doi.org/10.1109/ICCV.2007.4409052

Ramos Castro, D. 2007. *Forensic evaluation of the evidence using automatic speaker recognition systems*. Doctoral dissertation, Autonomous University of Madrid.

Reynolds, D.A. 1994. Speaker identification and verification using Gaussian mixture speaker models. In *Proceedings of the ESCA Workshop on Automatic Speaker Recognition, Identification, and Verification*, pp. 27–30.

Reynolds, D.A., Quatieri, T.F., and Dunn, R.B. 2000. Speaker verification using adapted Gaussian mixture models. *Digital Signal Processing*, 10: 19–41. https://doi.org/10.1006/dspr.1999.0361

Sadjadi, S.O. and Hansen, J.H.L. 2013. Unsupervised speech activity detection using voicing measures and perceptual spectral flux. *IEEE Signal Processing Letters*, 20(3): 197–200. https://doi.org/10.1109/LSP.2013.2237903

Silva, D.G. and Medina, C.A. 2017. Evaluation of MSR Identity Toolbox under conditions reflecting those of a real forensic case (forensic_eval_01). *Speech Communication*, 94: 42–49. http://dx.doi.org/10.1016/j.specom.2017.09.001

**Legal references**

## 13     Figure captions

**Figure 1.** Procedure for the calculation of MFCCs. The numbers in black circles correspond to the numbered steps in the main text. DFT = discrete Fourier transform. DCT = discrete cosine transform.

**Figure 2.** Feature warping. The original feature value is replaced by the warped feature value. The warping is achieved by mapping from the empirical cumulative probability distribution of the original feature values to a parametric target cumulative probability distribution, in this case the standard cumulative Gaussian distribution.

**Figure 3.** Examples of the effects of channel and feature warping on the distribution of MFCCs. The first column represents MFCC values extracted from a high-quality audio signal (the same in both rows). The next three columns represent the results of sequentially applying various signal processing techniques to simulate casework conditions. The top row represents a simulated questioned-speaker recording condition and the bottom row represents a simulated known-speaker recording condition. The final column represents the results of applying feature warping to the values represented in the immediately preceding column, applied separately to the simulated questioned-speaker recording condition and to the simulated known-speaker recording condition.

**Figure 4.** Example of using the EM algorithm to train a two-component GMM. The example is based on artificial data that were created for illustrative purposes. The dotted curve represents the distribution that was used to generate the data. The dashed curve represents the fitted GMM, and the black curve and the white curve represent the two Gaussian components of the fitted GMM. The top panel shows the initial GMM distribution based on a random seed. The data points (the circles on the $x$ axis) are shaded according to their responsibilities with respect to the two components. The second panel shows the fitted GMM distribution after 1 iteration of maximization (along with the responsibilities after that iteration). The third panel shows the results after 20 iterations, and the bottom panel the results after 40 iterations.

**Figure 5.** Example of using a GMM-UBM model to calculate a likelihood ratio for a single questioned-speaker-recording feature vector. This example is based on artificial two-dimensional data generated for illustrative purposes. The UBM is represented by the surface drawn with the darker mesh and the MAP adapted known-speaker GMM is represented by the surface drawn with the lighter mesh. The *x-y* location of the vertical line indicates a questioned-speaker-recording feature vector value at which the likelihoods of the UBM and the known-speaker GMM are being evaluated (the likelihoods are given by the *z* values of the intersections of the vertical line with each of the surfaces).

**Figure 6.** Example of using PCA to reduce the data from two dimensions to one dimension. The example uses artificial data created for illustrative purposes. Different shaped symbols represent different speakers and the two different intensities of shading represent two different conditions, e.g., a questioned-speaker-recording condition and a known-speaker-recording condition. The PCA dimension (represented by the oblique axis) is in the direction of maximum variance in the original *x-y* space.

**Figure 7.** Example of using CLDF for mismatch compensation and to reduce the data from two dimensions to one dimension. The example uses artificial data created for illustrative purposes (the same data as used for Figures 6). Different shaped symbols represent different speakers and the two different intensities of shading represent two different conditions, e.g., a questioned-speaker-recording condition and a known-speaker-recording condition. The CLDF dimension (represented by the oblique axis) is in the direction in the original *x-y* space that has the maximum ratio of between- versus within-speaker variance.

**Figure 8.** Example of using a PLDA model to calculate a score for a pair of i-vectors. The numerator of the PLDA model is represented by the surface drawn with the lighter mesh and the denominator is represented by the surface drawn with the darker mesh. The *x-y* location of the vertical line indicates the values of the i-vectors $v_q$ and $v_k$ (extracted from the questioned- and known -speaker recordings respectively) at which the likelihoods of numerator and denominator of the PLDA model are being evaluated (the likelihoods are given by the *z* values of the intersections of the vertical line with each of the surfaces).

**Figure 9.** Simplified example of the architecture of a feed-forward DNN consisting of an input layer, two hidden layers, and an output layer. Not all connection between nodes have been drawn, and weights have been indicated for only a few connections.

**Figure 10.** Illustration of the architecture of a DNN embedding system. Only the time-dimension is shown for the frame level (the frequency dimension is not shown). The input layer (at the bottom) spans *t*–7 through *t*+7. The time dimension is collapsed by the third hidden layer of the frame level. The statistics-pooling layer calculates the mean and standard deviation of each node in the preceding layer when the whole of the speech of interest in a recording is passed by the input layer. The frequency dimension is shown for the segment level.

**Figure 11.** Example of using linear discriminant analysis or logistic regression to convert a score to a likelihood ratio. The example uses artificial data created for illustrative purposes. Different-speaker training scores are shown as grey triangles and same-speaker training scores are shown as white circles. The top panel shows a linear discriminant analysis model fitted to the data. The middle and bottom panels can be derived from the top panel. The middle and bottom panels can also be derived by fitting a logistic regression model to the same data. The vertical line represents a score value that is being converted to a likelihood-ratio value. In the top panel the output is clearly the ratio of two likelihoods. The middle panel shows that the conversion from scores to log-likelihood-ratio values is a linear function. (In reality the plotted training data are illustrative only and the plotted functions show ideal values based on specified parametric distributions.)

**Figure 12.** Penalty functions for calculating $C_{llr}$. The same-speaker and different-speaker curves correspond to the functions within Equation 20's left and right summations respectively.

**Figure 13.** Examples of Tippett plots. The three plots represent three different systems tested on the same test data. The examples are based on artificial data created for illustrative purposes. The data represent 50 same-speaker test pairs and 200 different-speaker test pairs – imbalance in the number of same-speaker and different-speaker test pairs is usual since it is easier to construct different-speaker pairs than same-speaker pairs. The $C_{\text{llr}}$ values corresponding to the results shown in the top, middle, and bottom panels are 1.068, 0.698, and 0.307 respectively.

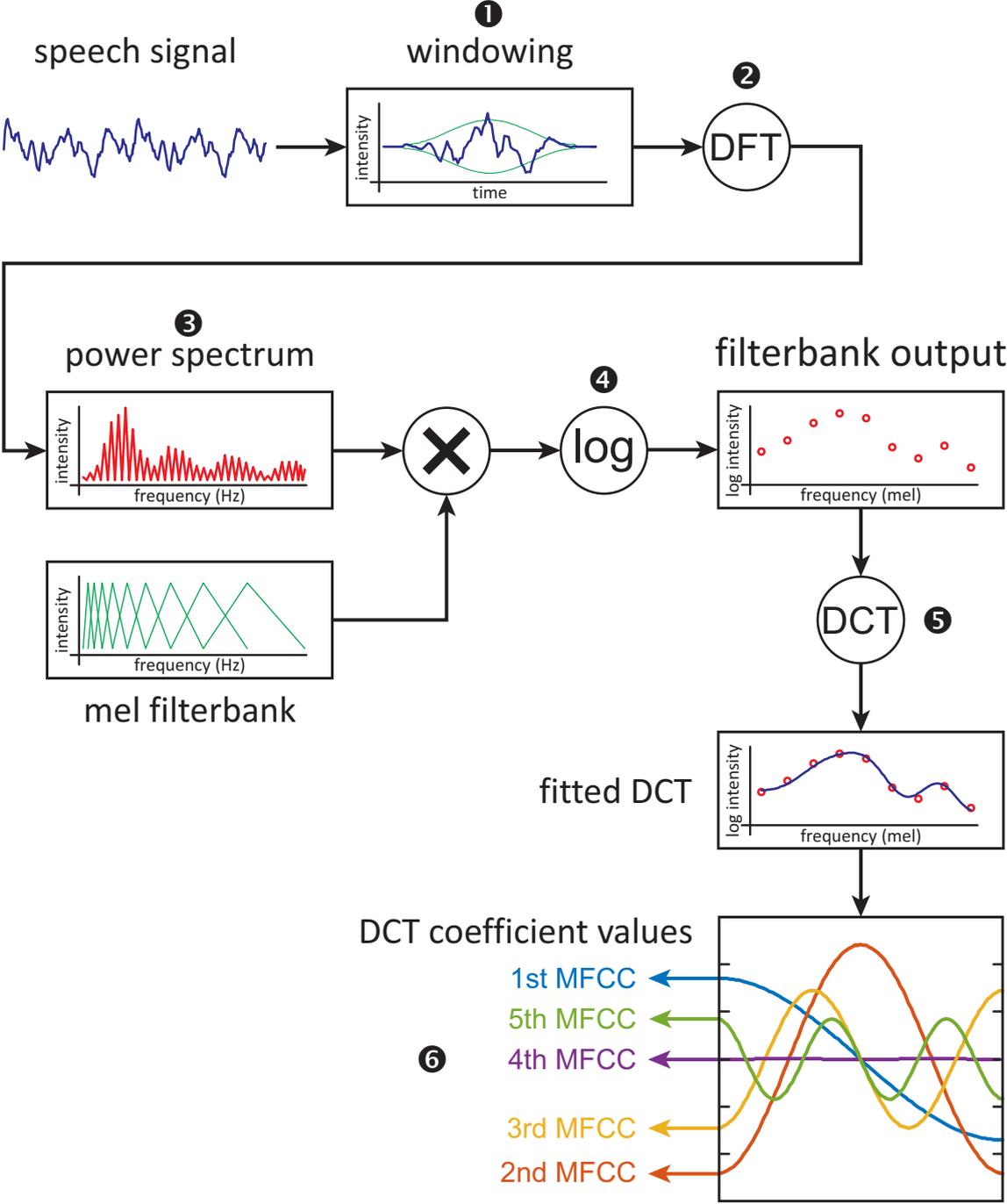

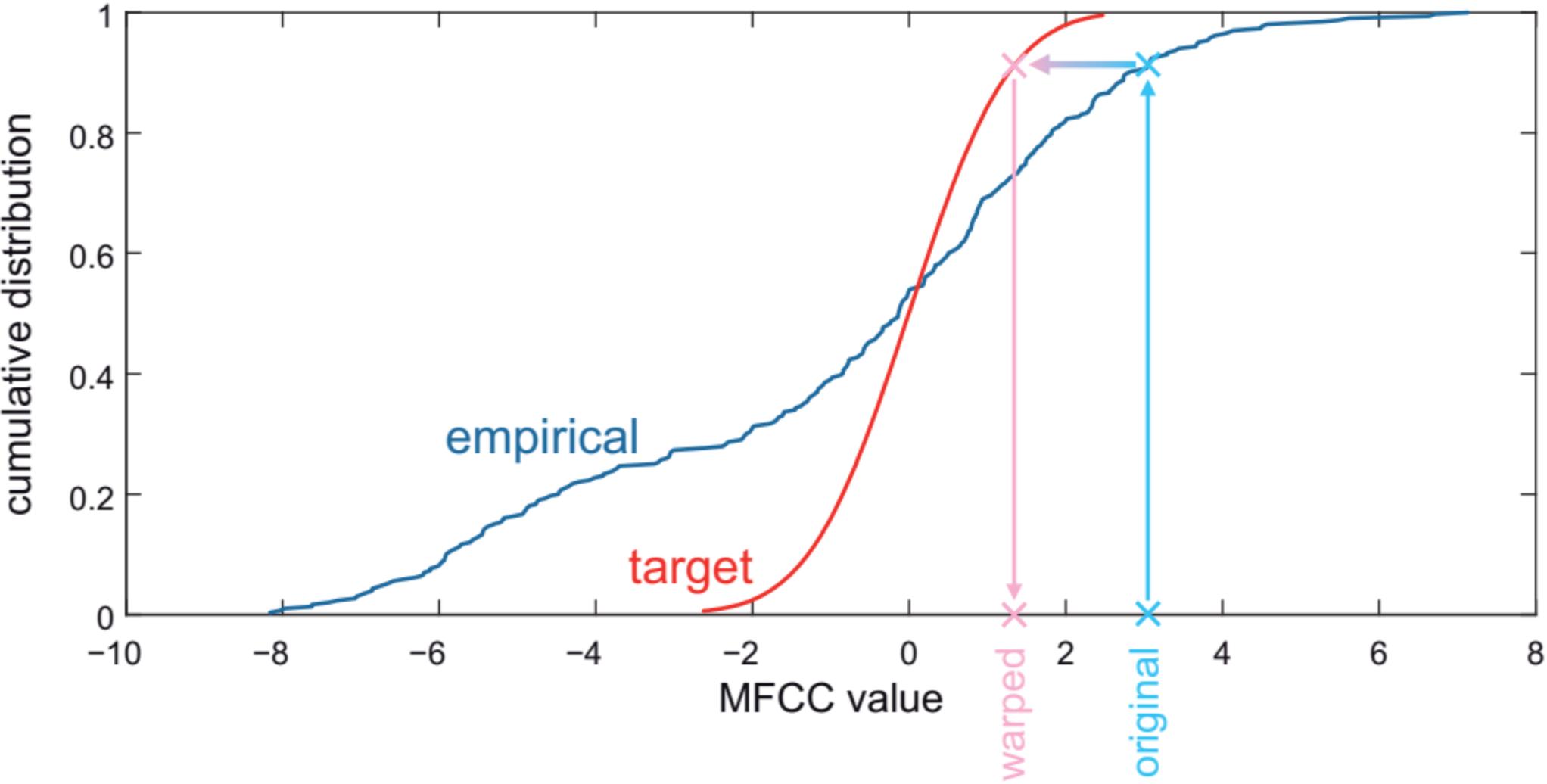

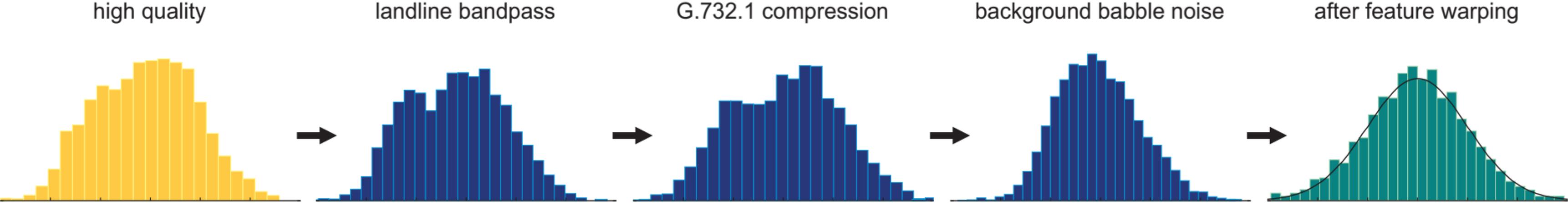
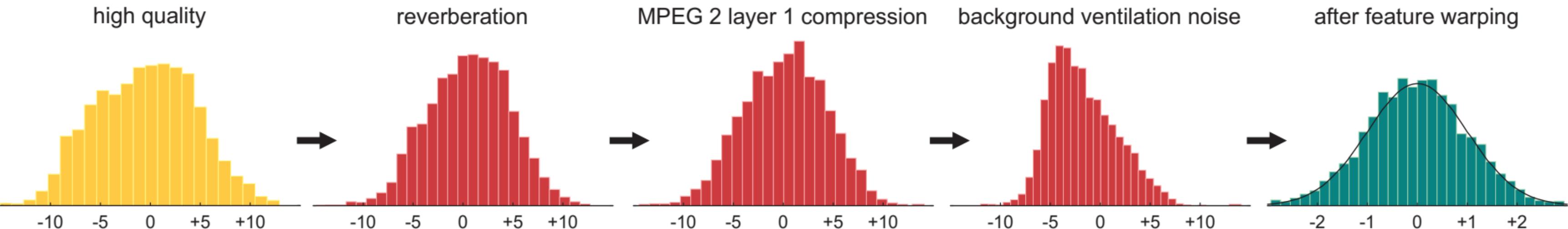

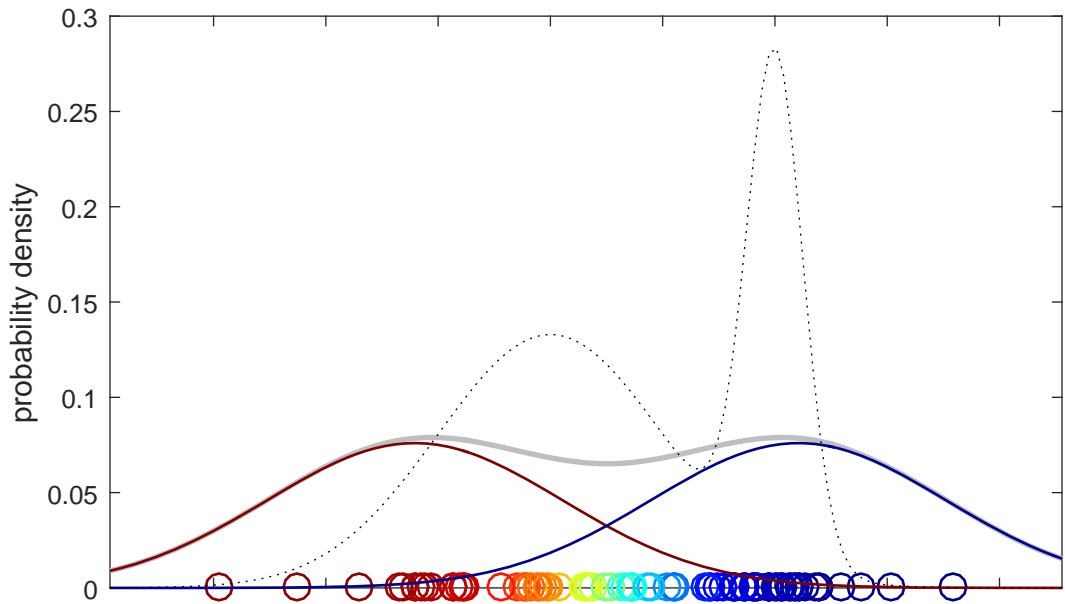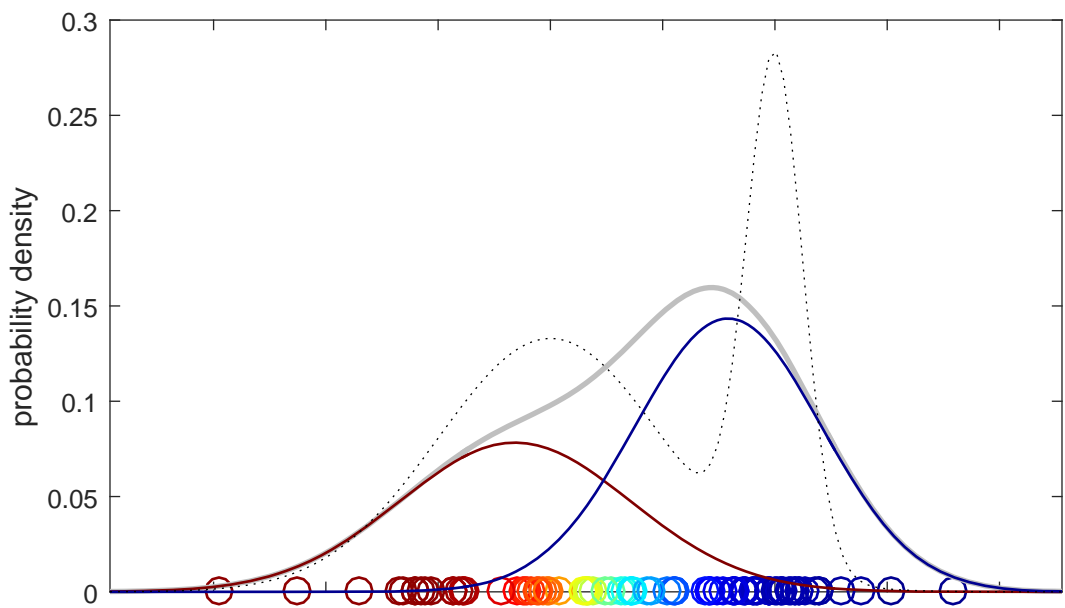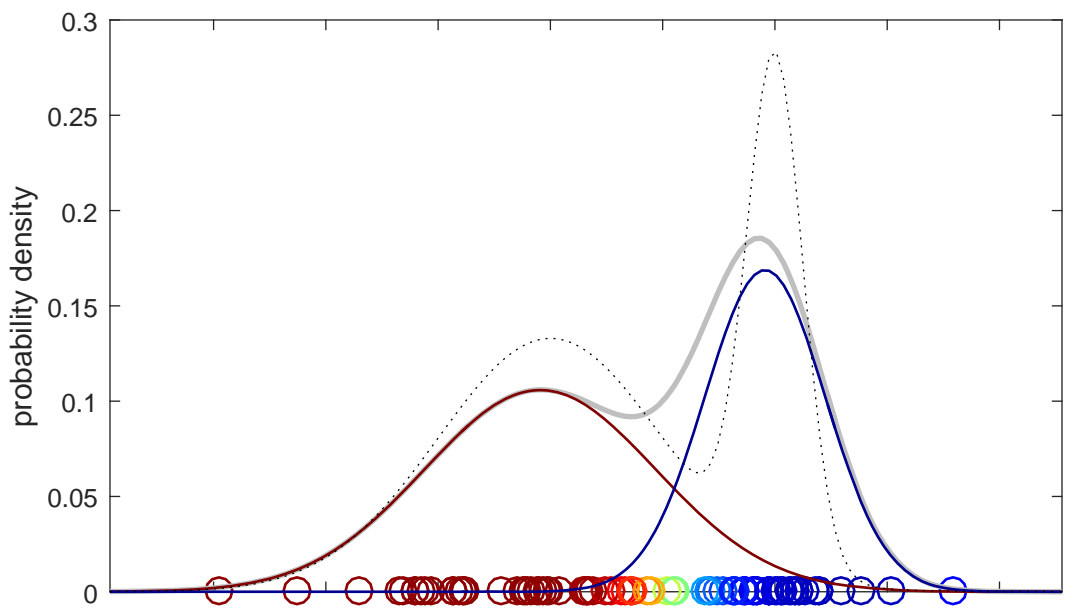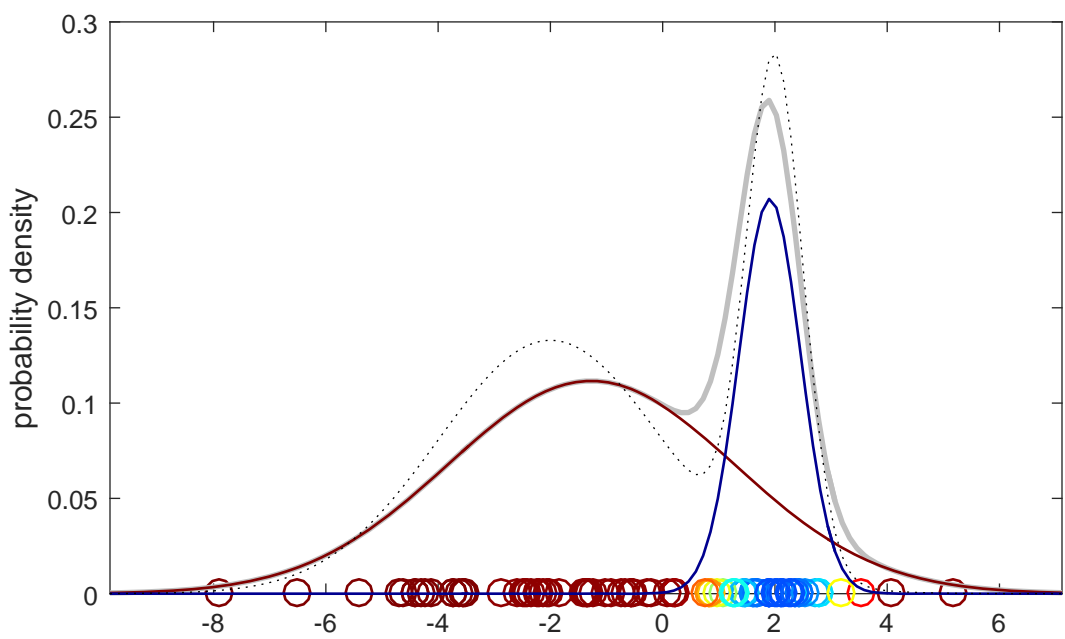

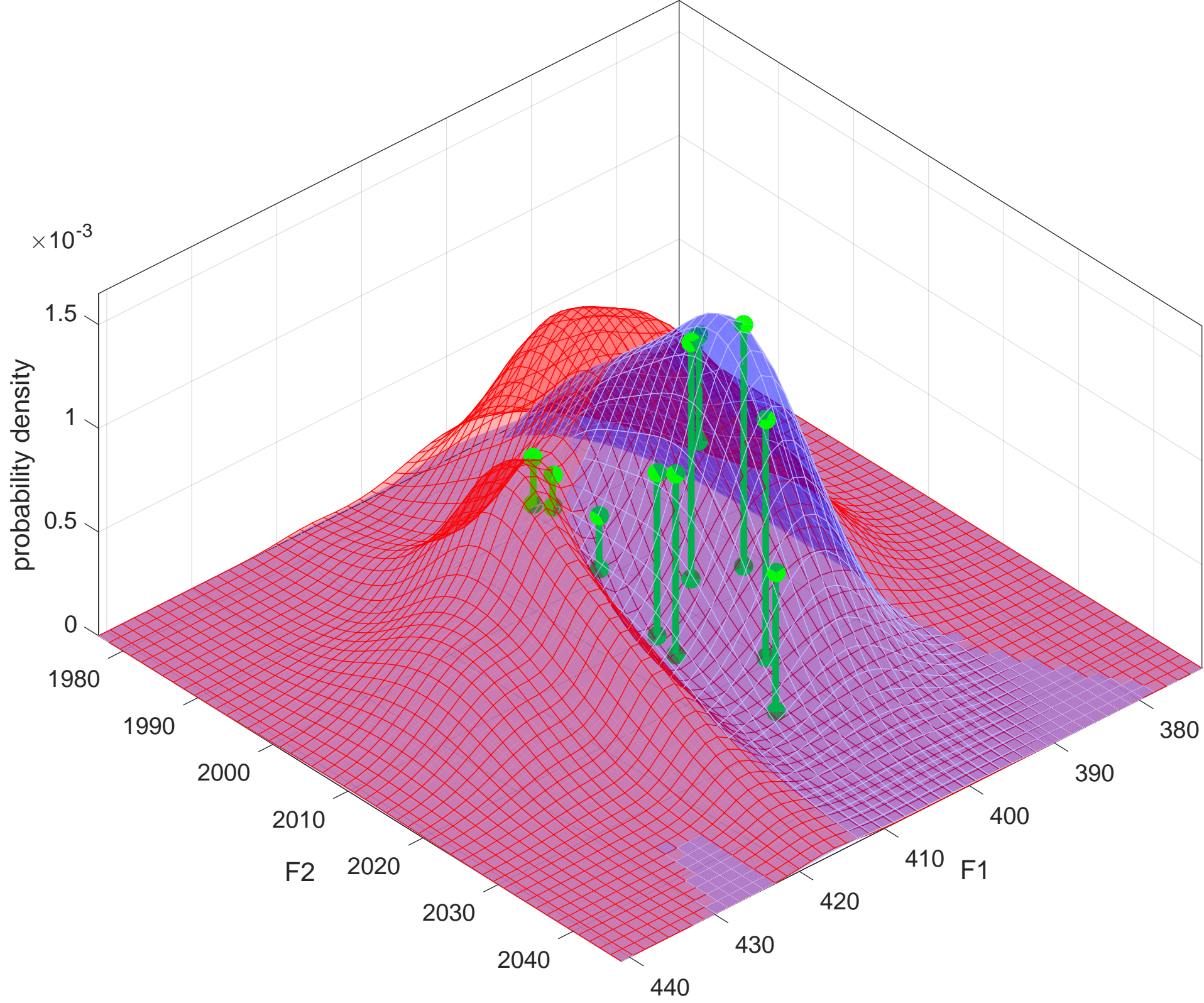

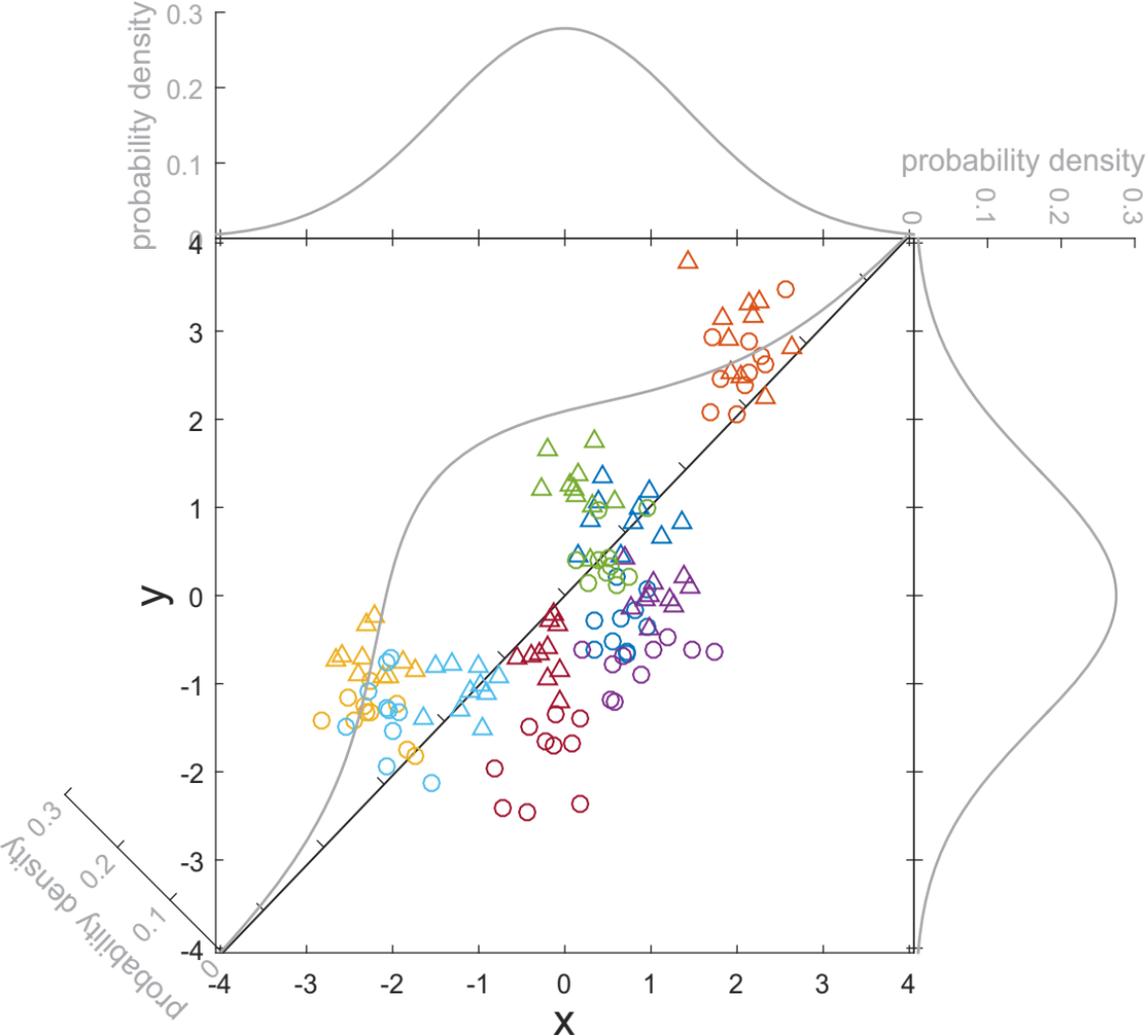

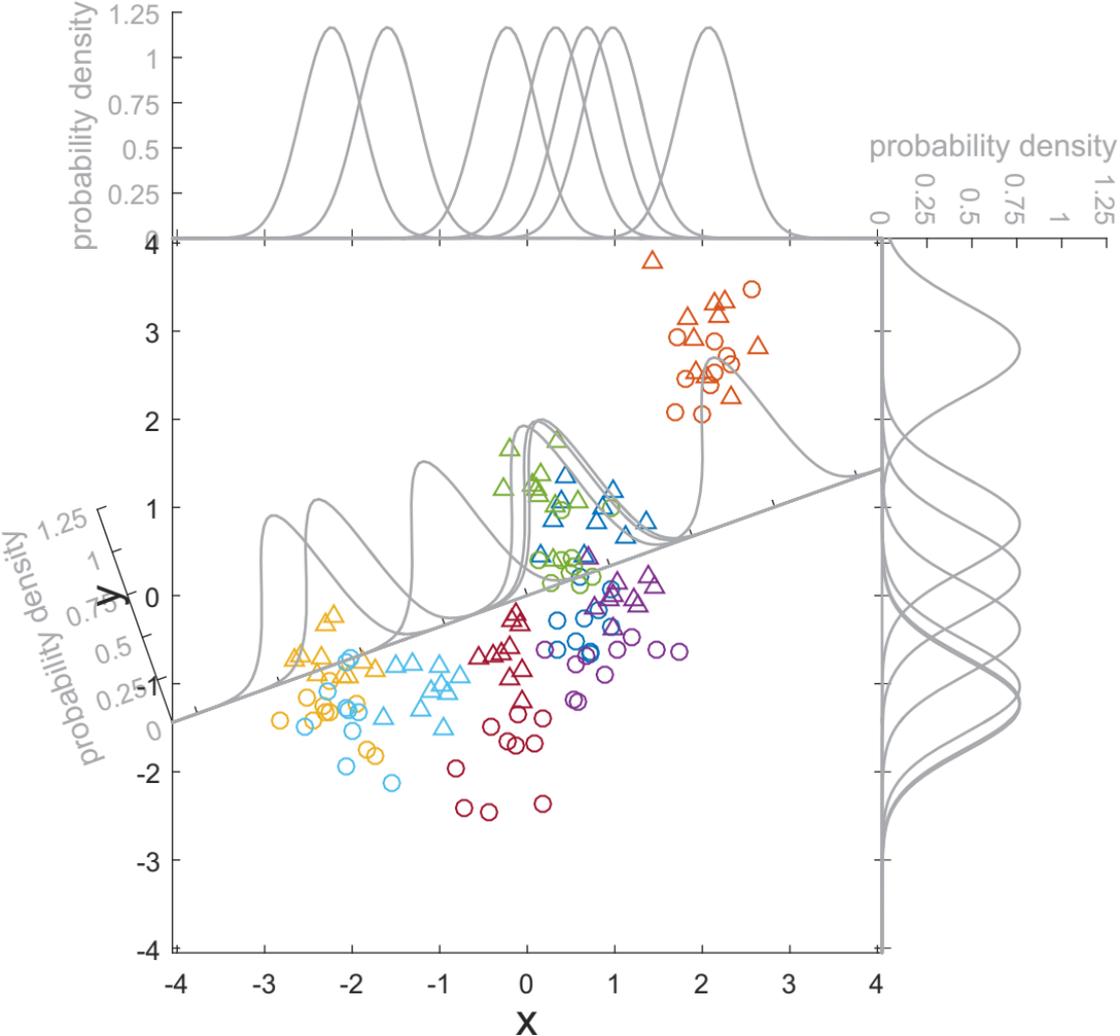

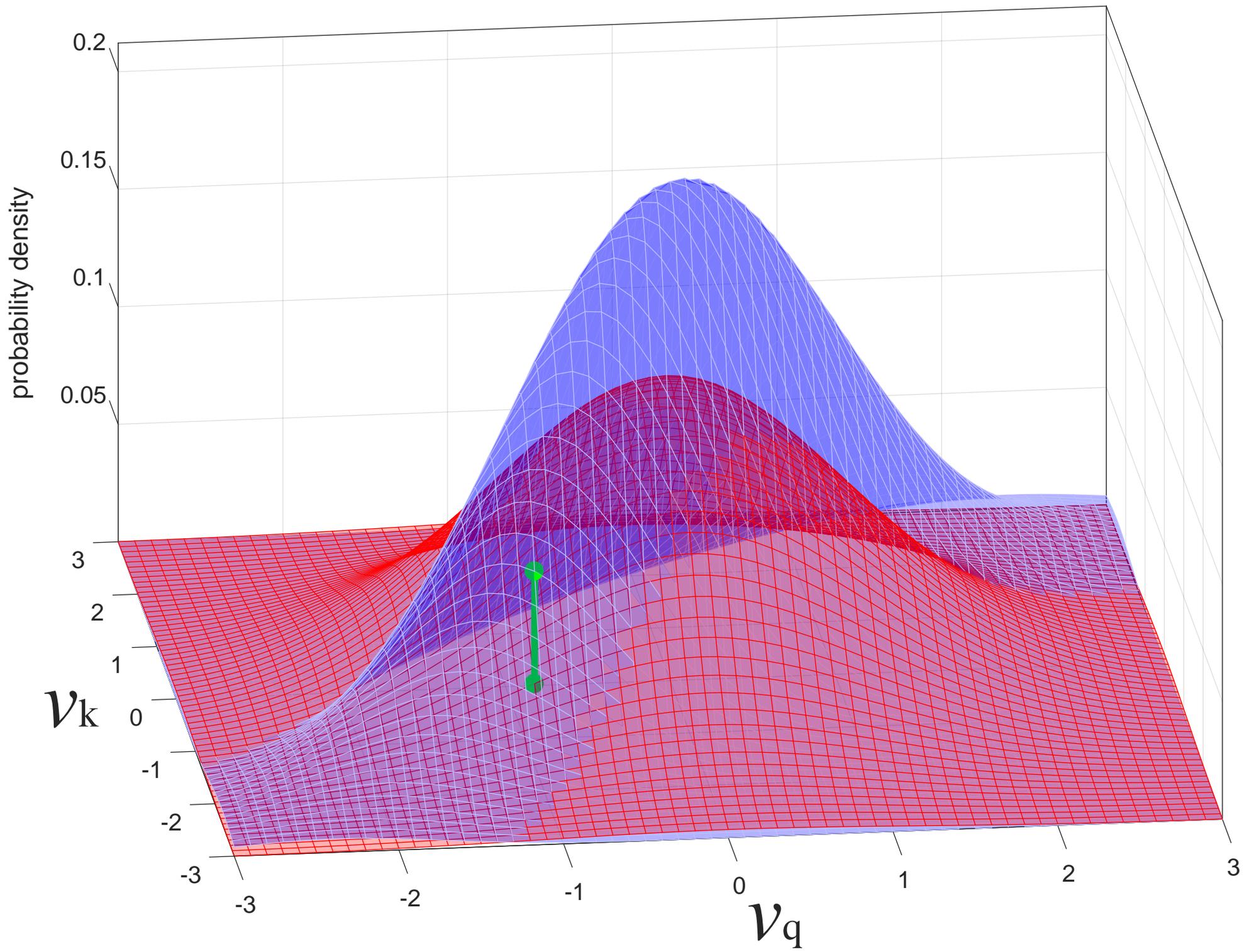

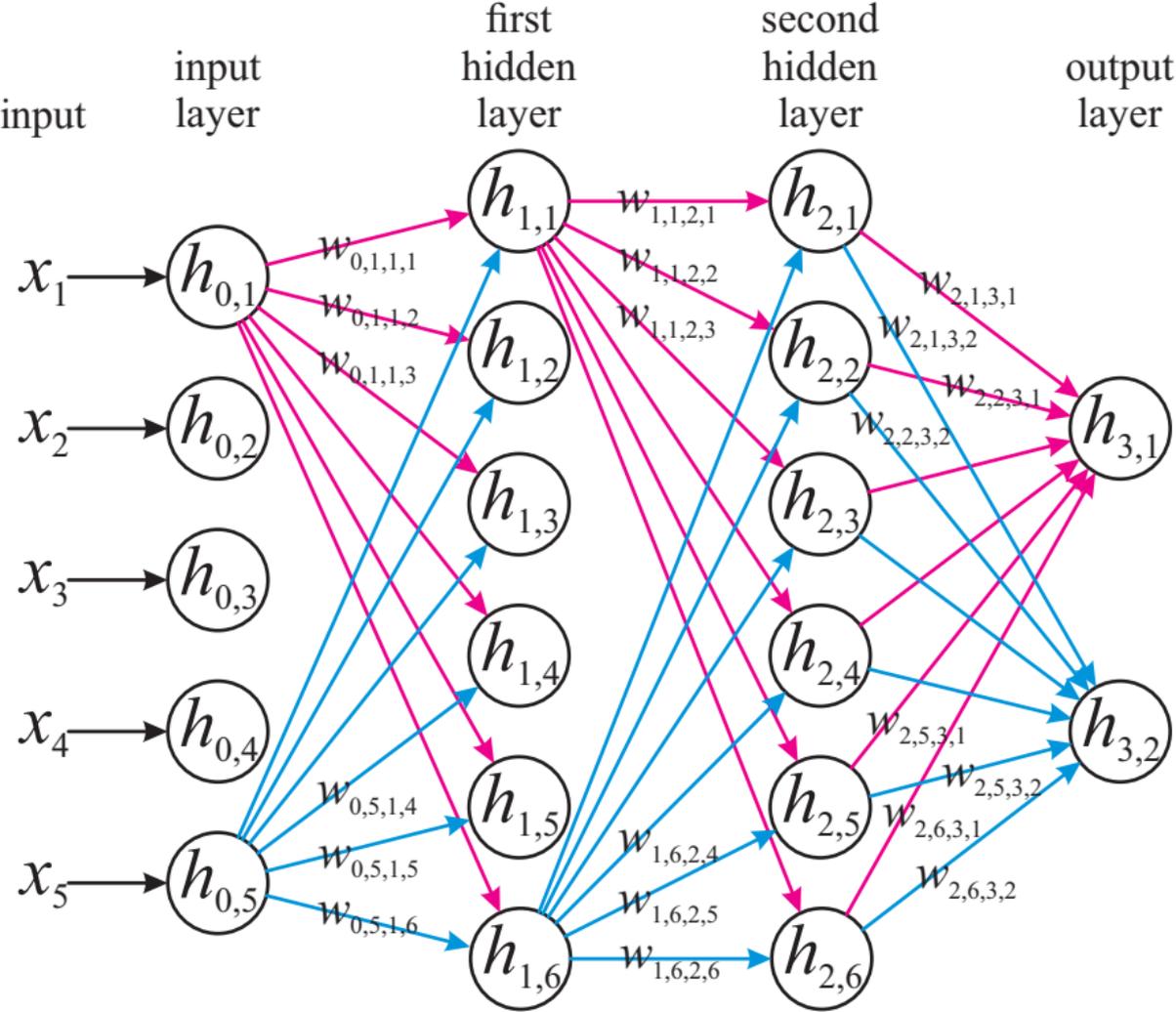

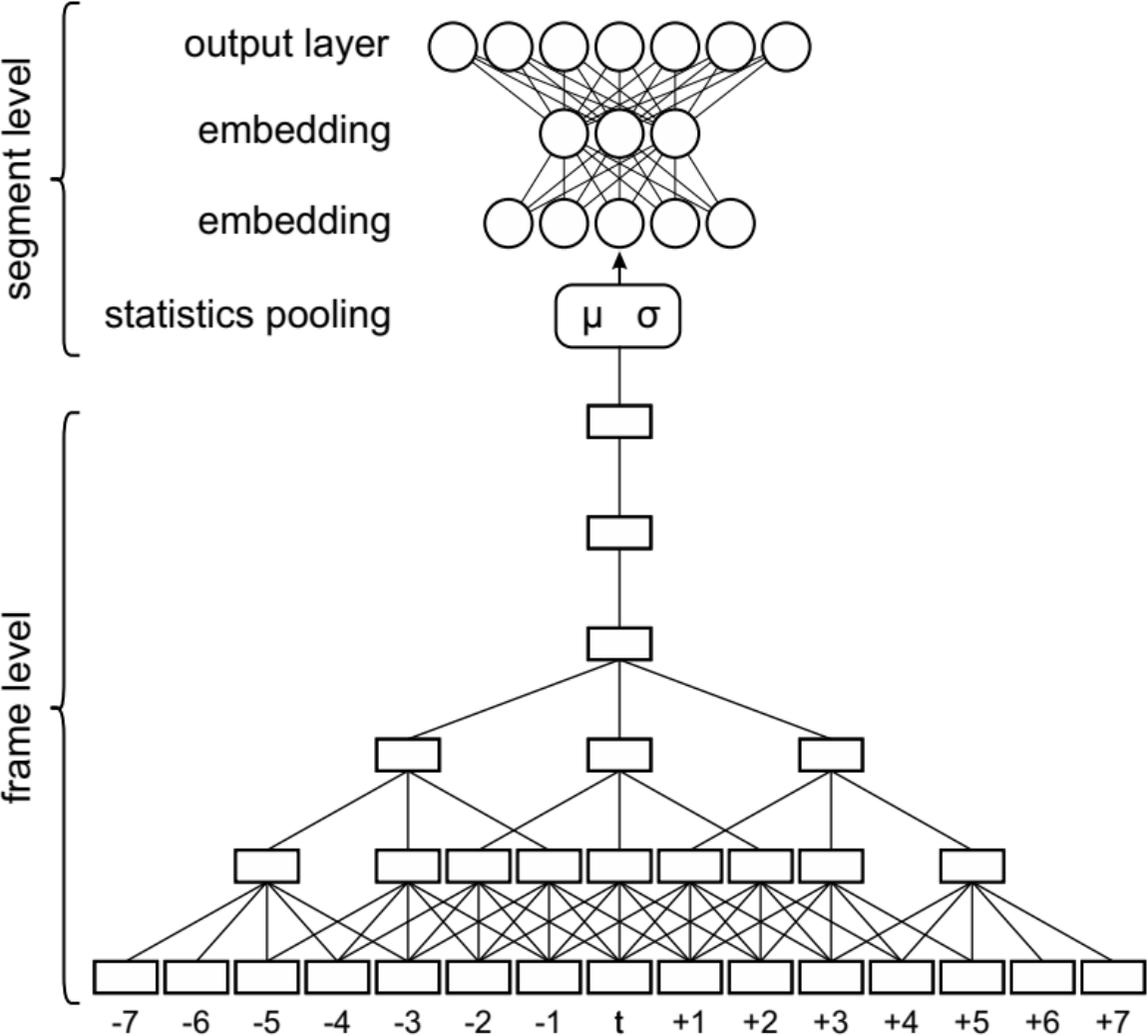

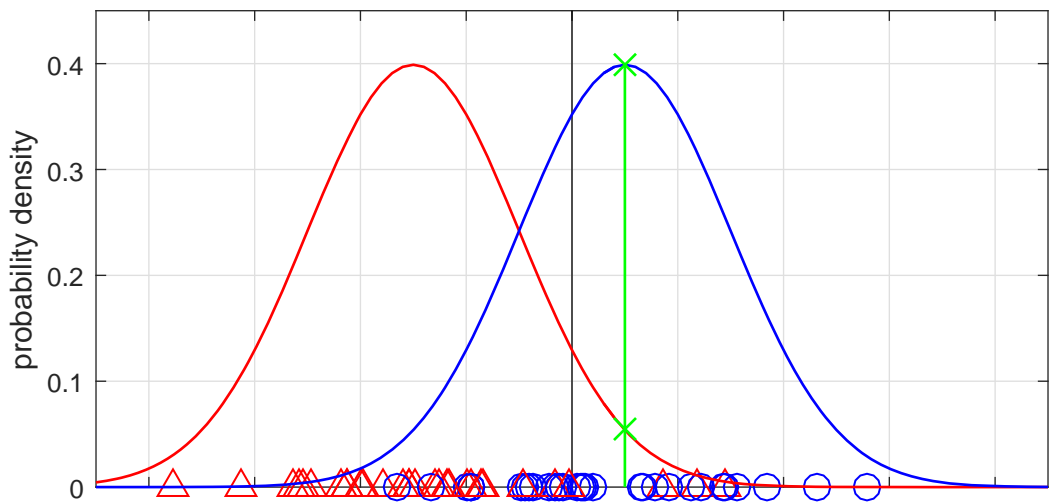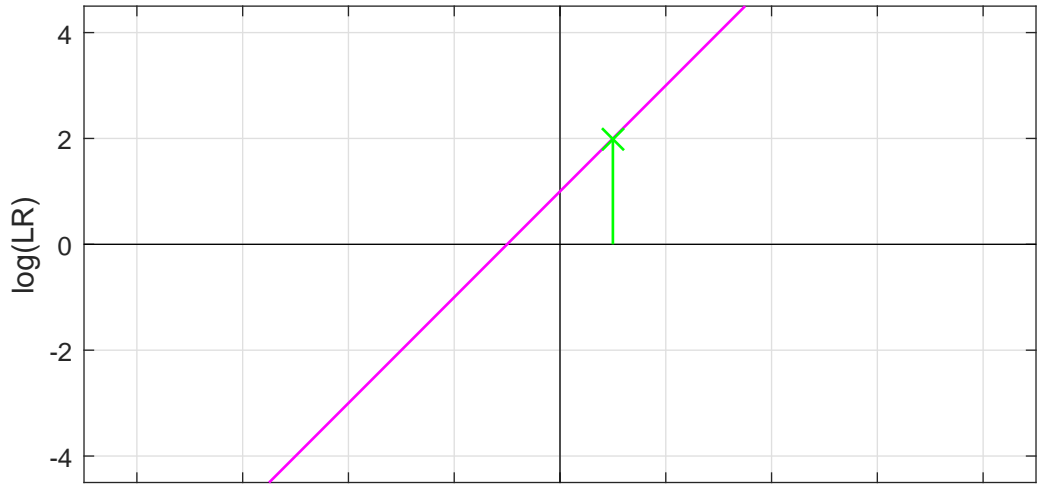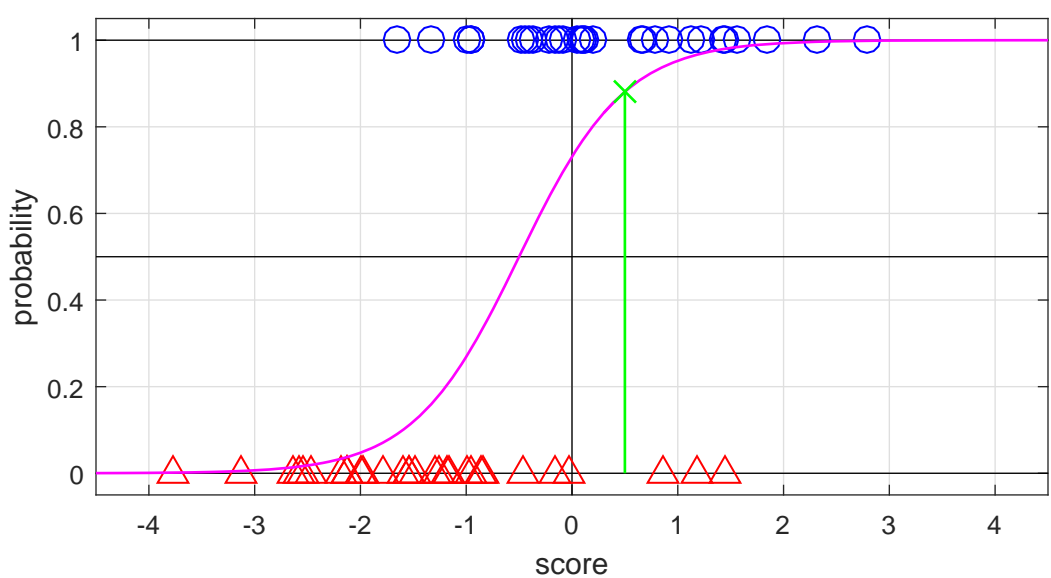

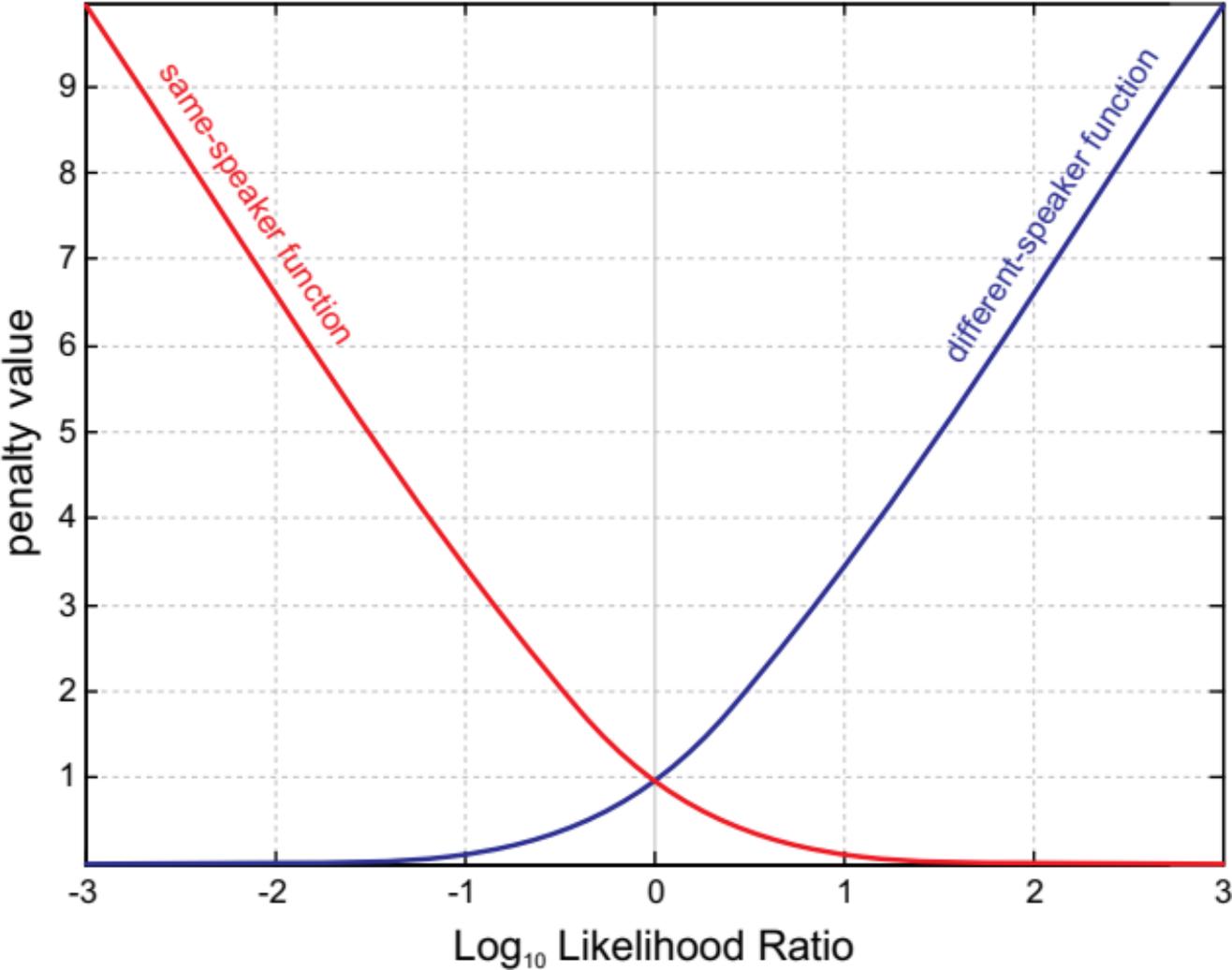

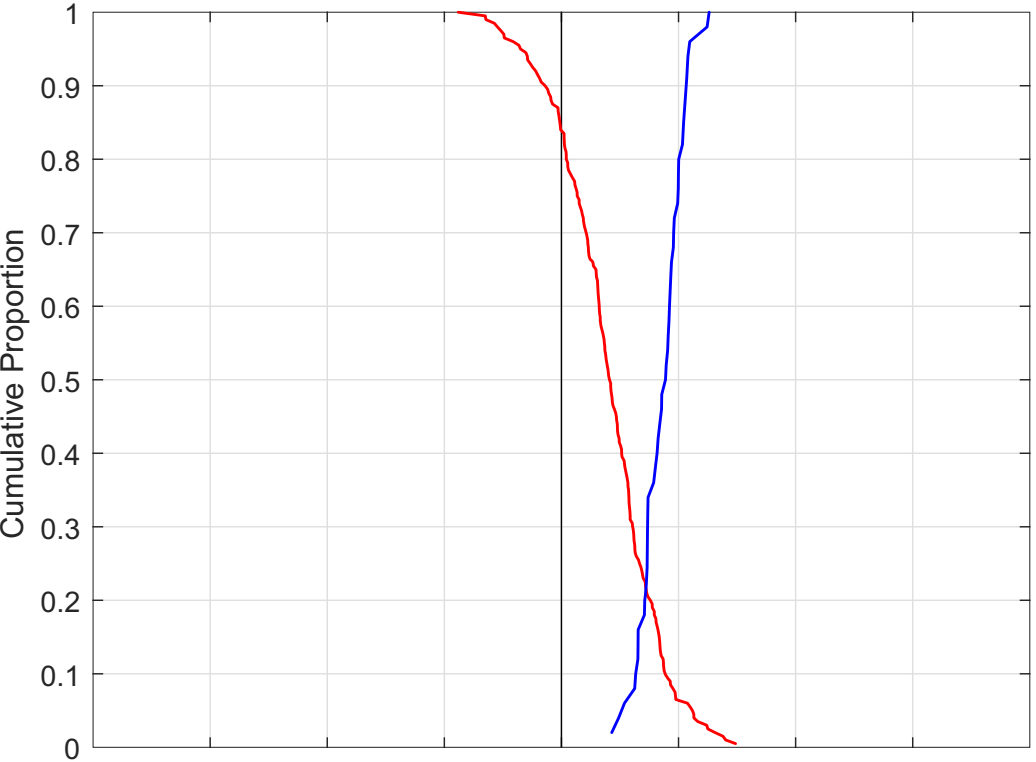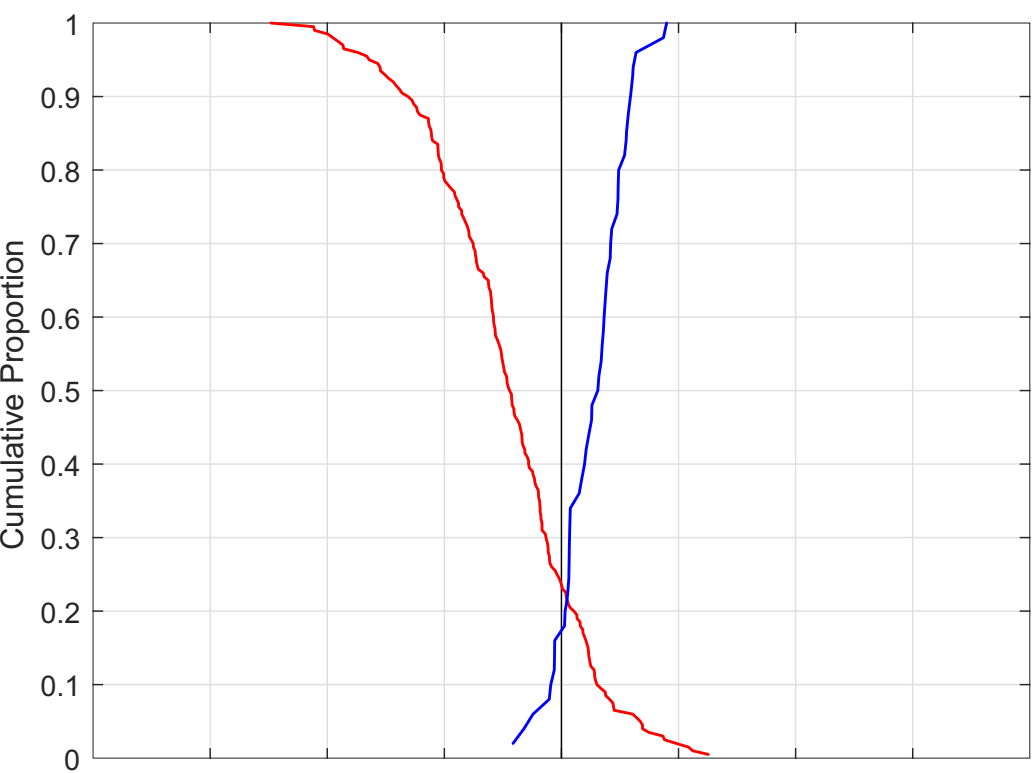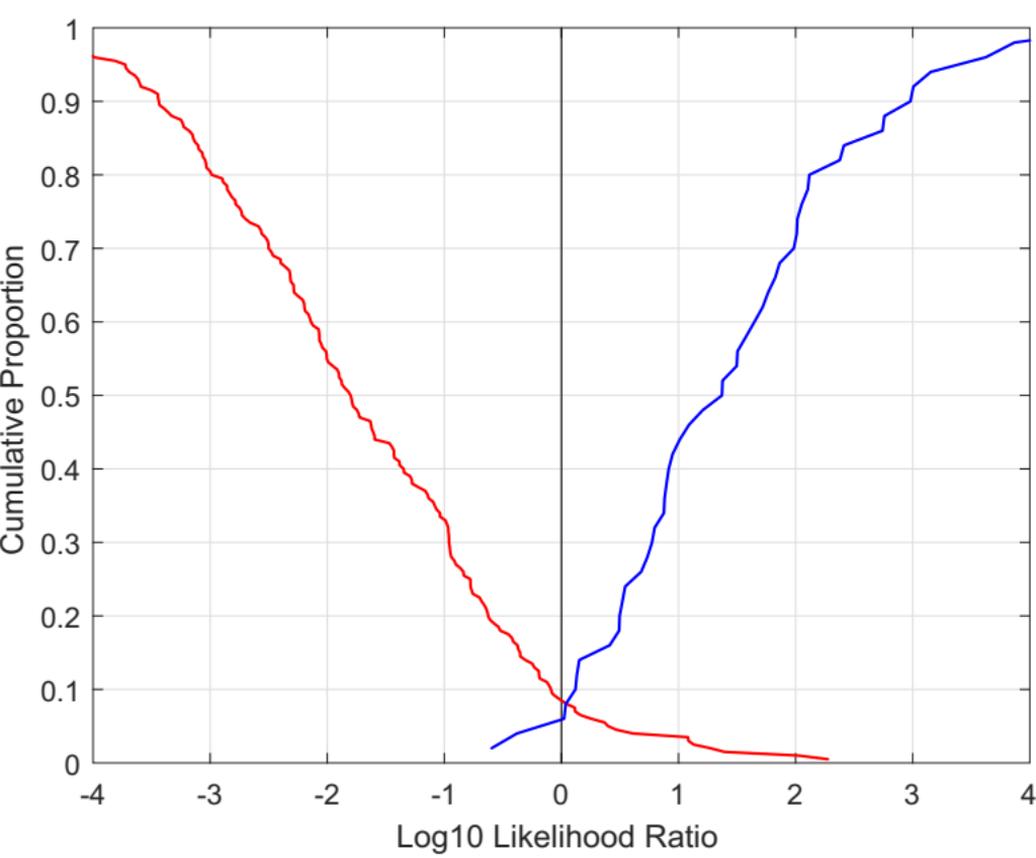